\documentclass[useAMS,usenatbib,usegraphicx]{mn2e}
\usepackage{amsmath}
\usepackage{graphicx}
\usepackage{dcolumn}
\usepackage{bm}
\usepackage{natbib}
\usepackage{multirow}
\usepackage{epsfig}
\usepackage{amssymb}
\usepackage{float}

\def\simlt{\lower.5ex\hbox{$\; \buildrel < \over \sim \;$}}
\def\simgt{\lower.5ex\hbox{$\; \buildrel > \over \sim \;$}}

\def\beq{\begin{equation}}
\def\eeq{\end{equation}}

\newcommand{\beqa}{\begin{eqnarray}}
\newcommand{\eeqa}{\end{eqnarray}}

\newcommand{\rmd}{{\rm d}}
\newcommand{\rme}{{\rm e}}

\renewcommand{\vec}[1]{{\bf #1}}
\newcommand{\rmi}{{\rm i}}

\newcommand{\pra}{{Phys.~Rev.~A~}}

\newcommand{\prl}{{Phys.Rev.Letters}}

\title[Hydrogen-proton collisions]{Excitation and charge transfer in H-H$^+$ collisions at 5--80 keV and application to astrophysical shocks}

\date{\today}

\author[Tseliakhovich, Hirata \& Heng]
{Dmitriy Tseliakhovich$^{1}$\thanks{E-mail: dimlyus@caltech.edu},
Christopher M. Hirata$^2$ and Kevin Heng$^3$\\$^{1}$California
Institute of Technology, M/C 249-17, Pasadena, California 91125, USA\\$^{2}$California
Institute of Technology, M/C 350-17, Pasadena, California 91125, USA\\$^{3}$ETH Zurich, Institute for Astronomy, Wolfgang-Pauli-Strasse 27, CH-8093, Zurich, Switzerland}

\begin{document}

\maketitle

\begin{abstract}
In astrophysical regimes where the collisional excitation of hydrogen atoms is relevant, the cross sections for the interactions of hydrogen atoms with electrons and protons are necessary for calculating line profiles and intensities.  In particular, at relative velocities exceeding $\sim 1000$ km s$^{-1}$, collisional excitation by protons dominates over that by electrons.  Surprisingly, the H-H$^+$ cross sections at these velocities do not exist for atomic levels of $n \ge 4$, forcing researchers to utilize extrapolation via inaccurate scaling laws.  In this study, we present a faster and improved algorithm for computing cross sections for the H-H$^+$ collisional system, including excitation and charge transfer to the $n \ge 2$ levels of the hydrogen atom.  We develop a code named \texttt{BDSCx} which directly solves the Schr\"{o}dinger equation with variable (but non-adaptive) resolution and utilizes a hybrid spatial-Fourier grid.  Our novel hybrid grid reduces the number of grid points needed from $\sim 4000 n^6$ (for a ``brute force", Cartesian grid) to $\sim 2000 n^4$ and speeds up the computation by a factor $\sim 50$ for calculations going up to $n=4$. We present $(l,m)$-resolved results for charge-transfer and excitation final states for $n = 2$--4 and for projectile energies of 5--80 keV, as well as fitting functions for the cross sections.  The ability to accurately compute H-H$^+$ cross sections to $n=4$ allows us to calculate the Balmer decrement, the ratio of H$\alpha$ to H$\beta$ line intensities. We find that the Balmer decrement starts to increase beyond its largely constant value of 2--3 below 10 keV, reaching values of 4--5 at 5 keV, thus complicating its use as a diagnostic of dust extinction when fast ($\sim 1000$ km s$^{-1}$) shocks are impinging upon the ambient interstellar medium. 
\end{abstract}


\section{Introduction}
The extreme densities (both low and high) and high temperatures inherent in many astrophysical processes allow for atomic interactions beyond the realm of terrestrial consideration or experiments.  Hydrogen, the most abundant element in the Universe, signals its presence through the production of emission (or absorption) lines, either via the recombination of protons with electrons or the collisional excitation of hydrogen atoms by electrons or protons.  When the medium is tenuous ($\sim 1$ cm$^{-3}$) and the relative velocity of interaction is high ($\sim 1000$ km s$^{-1}$), recombination becomes slow and collisional excitation dominates.

Fast, astrophysical shocks are examples where collisional excitation becomes important.  In particular, a class of shocks known as ``Balmer-dominated shocks" are driven by astrophysical pistons (e.g., supernova remnants, pulsar wind nebulae, novae) impinging upon the ambient interstellar medium, producing hydrogen lines observed in the Lyman and Balmer series accompanied by a dearth of metal lines (see \citealt{Heng10} for a review).  Both the processes of excitation and charge transfer are inferred to be at work in these Balmer-dominated shocks, which may also be relevant in young, high-redshift galaxies \citep{hs08}.  The line widths and intensities serve as diagnostics for the shock velocities and temperatures, which necessitates the knowledge of excitation cross sections to hydrogen levels $n \geq 4$, especially if one is interested in emission lines such as H$\beta$ as well as the Paschen and Brackett lines.  The ratio of H$\alpha$ to H$\beta$ line intensities, known as the ``Balmer decrement", further serves as a diagnostic for dust extinction due to its insensitivity to electron temperature and density \citep{draine11} --- it is of interest to establish if this insensitivity extends to the regime occupied by fast, astrophysical shocks.

At relative velocities of $\sim 1000$ km s$^{-1}$ or greater, the excitation of hydrogen atoms becomes dominated by the interactions with protons rather than electrons \citep{hm07}.  A glaring and surprising gap in the existing literature, both from theory or experiment, is that cross sections for the reaction,
\begin{equation}
\mbox{H(1s)} + \mbox{H}^+ \rightarrow \mbox{H}\left(n\ge4\right) + \mbox{H}^+,
\end{equation}
are essentially non-existent at these velocities, corresponding to energies of about 5 keV.  In the absence of these cross sections, some researchers have resorted to using approximate scaling laws such as 
\begin{equation}
\sigma_{nl} = \left( \frac{n_0}{n} \right)^{3} \sigma_{n_0 l},
\label{eq:scaling}
\end{equation}
where $n_0 < n$, to extrapolate for cross sections with $n \ge 4$ using available ones with $n_0 \le 3$. This scaling law can be derived using the Born approximation and is only approximately valid at high velocities $v \gg \alpha c$, or $E_{col} \gg 25$ keV \citep{Salin88}. Here by velocity we mean relative velocity between colliding particles, $E_{col}$ is the kinetic energy of a hydrogen atom moving towards the proton at rest. Besides the inaccuracy associated with extrapolation, it also leaves open the question of how to obtain cross sections for levels with $l$ values which do not exist for $n_0 \le 3$ (e.g., 4f).

Initial attempts to calculate and measure cross sections of hydrogen collisions date back to the 1960s \citep{Stebbings65,Wilets66,Ryding66,Bayfield69}.  On the theoretical side, the success of these efforts has been hindered by the high computational cost of numerical simulations.  On the experimental side, it has been limited by stringent requirements of creating high vacuum states and the high costs of preparing and characterizing atomic hydrogen targets. 

In calculating cross sections for high-$nl$ proton-hydrogen collisions, it is important to consider several distinct cases. At low velocities ($v \ll \alpha c$), collisions between a hydrogen atom and a proton lead to large deflections of the colliding particles and, as in the case of an unbound $H_2^+$ molecule, the electron wave function deforms adiabatically during the collisional time. In such situations, the initial configuration of the system is considerably modified during the collision and the process must be treated in a way which reflects the interplay between various quantum states of the electronic wave function. This problem has been addressed with the close-coupling approximation which assumes that, during the atomic collision, the electron wave function transitions between a certain number of configurations which form the ``basis set'' of functions \citep{Fritsch91}. The dominant outcome of the low energy collisions is charge exchange/transfer between the colliding particles. 

At high energies ($v \gg \alpha c$), which includes the relativistic regime, colliding particles follow undeflected, straight-line trajectories. This case is well described by the Born approximation in which the incoming proton is seen as a small perturbation of the electronic wave function \citep{Bates53}. The dominant outcome of the collisions in the relativistic regime is the excitation or ionization of the hydrogen atom. 

The third case corresponds to intermediate velocities ($v \sim \alpha c$), where collisional times are of the order of the atomic timescale and therefore a perturbative treatment of the problem becomes invalid. This is precisely the regime relevant to Balmer-dominated shocks.  In this regime, the behaviour of the electron wave function is more complicated than at low energies or relativistic energies. At intermediate energies ($E \sim 10$ keV), there is no dominant outcome for a collision: charge transfer, collisional excitation and ionization are all important and interconnected. There is no clear intuitive picture of the mechanism for populating various electron quantum states and therefore the use of the close-coupling approximation is challenging and requires development of multiple basis sets and extensive convergence tests. Analysis of the proton-hydrogen collisions at intermediate energies in the close-coupling approximation is an active area of research \citep{Fritsch91, Ford93, Kuang96, McLaughlin97, Martin99, Toshima99, Winter09, Crothers92, Brown96}.  However, the accuracy of the obtained results is still not fully determined as convergence tests of these methods are extremely hard, especially if extended over a large range of energies \citep{Ford93, Kuang96}.

An alternative method to addressing the problem of proton-hydrogen collisions at intermediate energies is via the direct solution of the Schr\"odinger differential  equation on a numerical grid \citep{Maruhn79, Bottcher82, Kulander82}. In fact, the grid-based method can be thought of as a finite basis set method with one basis function for each point on the grid. Over the past decade, several groups have taken this approach \citep{Kolakowska98,Kolakowska99}, producing results for energies ranging from 10 to 100 keV.  It is easier to test the convergence properties of these numerical grid methods, but the price to pay is that they are are notoriously computationally demanding due to the long-range nature of the Coulomb electrostatic force. This is especially true if one needs to accurately represent states of high $n$; for this reason, previous results were limited to $n \leq 3$. 

In most cases, the results obtained in previous studies measure and/or calculate cross sections of hydrogen collisions in the velocity range $\sim 100$--1000 km s$^{-1}$ only with a precision $\sim 10$--30\% (see \citealt{Heng10} and references therein). There is also a substantial disagreement between experimental results and theoretical calculations (e.g., \citealt{Winter09, Sidky01}). In the case of final states with $n>3$, robust theoretical or experimental cross-sections in the energy range relevant to Balmer-dominated shocks studies do not exist at all.

The objectives of the present study may be concisely stated as follows:
\begin{itemize}

\item To introduce a novel hybrid grid (Figure~\ref{Fig:Grid}) for the direct solution of the Schr\"{o}dinger equation;

\item To demonstrate that the use of this grid reduces the number of grid points needed from $\sim 4000 n^6$ (for a ``brute force", Cartesian grid) to $\sim 2000 n^4$, which corresponds to a gain in the speed of computation by a factor $\sim 50$ for $n \leq 4$ case; 

\item To provide cross sections for excitation and charge transfer reactions, in the H-H$^+$ collisional system up to $n = 4$, at energies of 5--80 keV;

\item To provide fitting functions for these cross sections so as to enable their (convenient) use by astrophysicists and astronomers;

\item To quantify the error associated with using the scaling law from equation (\ref{eq:scaling});

\item To calculate the Balmer decrement in the regime where fast ($\sim 1000$  km s$^{-1}$) astrophysical shocks are impinging upon ambient interstellar medium ($\sim 1$ cm$^{-3}$).

\end{itemize}

The rest of the paper is organized as follows. In Section~\ref{Sec:Model}, we provide a detailed description of the theoretical model behind our analysis and the major constraints driving the development of the code for high-$nl$ cross section calculations. Section~\ref{Sec:Code} describes the code developed for our calculations and shows the results of extensive consistency tests. We discuss capabilities and limitations of the code and provide guidance on how this code can be expanded and used by other groups. In Section~\ref{Sec:Results}, as well as in Appendixes A and B, we provide results of our cross section calculations and compare our results with earlier studies. In Section~\ref{Sec:Applications}, we briefly discuss astrophysical applications of the obtained cross sections with a specific focus on Balmer-dominated shocks. Our results are summarized in Section~\ref{Sec:Summary}. 

\section{Computational model for cross section calculations}
\label{Sec:Model}

\subsection{Initial setup for precise cross section calculations}

Our objective is to determine the cross sections for reactions of the form:
\begin{equation}
{\rm H}_{\rm A}(1s) + {\rm H}_{\rm B}^+ \rightarrow X.
\label{eq:rxn1}
\end{equation}
The hydrogen nuclei are assumed to be very massive so that their motion can be treated classically and one only 
has to follow the evolution of the electron wave function in the potential created by the two nuclei.  Except for extremely small impact parameter $b\le 
m_p^{-1}v^{-2}$, all potential energies are negligible compared to the nuclear kinetic energy, so we may use an undeflected (straight line, constant velocity) 
trajectory for the nuclei.  In this limit, one can distinguish the two nuclei --- hence their description as H$_{\rm A}$ and H$_{\rm B}$.  

The initial electronic state is that of the $1s$ orbital of atom A, i.e., $|1s_{\rm A}\rangle$.  The final states $X$ under consideration correspond to (i) no 
reaction ($|1s_{\rm A}\rangle$); (ii) excitation ($|nlm_{\rm A}\rangle$, $n\ge 2$); (iii) charge transfer ($|nlm_{\rm B}\rangle$); and (iv) ionization (everything else).  All of these are of interest, even for high $n$ levels, for example for the H${\beta}$ lines the upper level is $n=4$ and for the Br${\alpha}$ lines the upper level is $n=5$.

We choose a coordinate system such that the relative velocity points along the $z$-axis, ${\bmath V}=V\hat{\bmath e}_z$, and the nuclear separation vector lies in the $xz$-plane.  The relative separation is
\begin{equation}
{\bmath r}_{\rm A}-{\bmath r}_B = b\hat{\bmath e}_x + Vt\hat{\bmath e}_z.
\label{eq:sep}
\end{equation}
Note that the electron wave function is always symmetric under reflection across the $xz$-plane.  We will choose the origin of the coordinate system  
in the $x$-direction such that $x_{\rm A}=\frac12b$, and $x_{\rm B}=-\frac12b$.  The choice of origin in the $z$-direction will be discussed later.

The cross section to produce a particular final state $X$ is given by
\begin{equation}
\sigma_X = \lim_{T\rightarrow\infty}\int_0^\infty 2\pi b \left| \langle X|\hat S(-T,T)|1s_{\rm A}\rangle \right|^2\,\rmd b,
\label{eq:sigma-X}
\end{equation}
where $\hat S(t_{\rm i},t_{\rm f})$ is the time evolution operator from time $t_{\rm i}$ to $t_{\rm f}$.  The $S$-matrix element can in principle be obtained by evolving $|1s_{\rm A}\rangle$ forward in time, $|X\rangle$ backward in time, or some combination of both. For example, one could evolve both states to $t=0$, that is one could factor $\hat S$ as $\hat S(-T,T)=\hat S(0,T)\hat S(-T,0)$ and have $\hat S(0,T)$ back-operate on $\langle X|$. The results in the present paper are based on evolving $|1s_{\rm A}\rangle$ forward as this is the most efficient way to generate cross sections for large numbers of final states.

Computation of the matrix elements requires us to solve the Schr\"odinger equation. In this work, we focus on grid methods because they allow for easier convergence tests and can be applied over a wide range of energies without significant modifications. Numerical grid methods also allow direct visualization of wave function evolution during the collision. 

Throughout this paper we use atomic units for all quantities, i.e., energy in hartrees, length in Bohr radii, velocity in units of $\alpha c \approx 2190$ km s$^{-1}$, and mass in electron masses. Conversions between the Syst\`eme Internationale (SI) units and atomic units are provided in Table~\ref{tab:AtomicUnits}. 

\begin{table*}
		\begin{tabular}{lcccccr}
		\hline
 Dimension & & Name & & Expression & & Value in SI \\
 \hline \hline 
 Length & & Bohr radius & & $a_0 = \hbar/(m_e c \alpha)$ & & $5.29\times 10^{-11} \ $m \\
 Energy & & Hartree & & $E_h = \alpha^2 m_e c^2$ & & $4.36\times 10^{-18} \ $J \\
 Velocity & & & & $\alpha c$ & & $2.19\times 10^{6} \ $m s$^{-1}$ \\
 Electric field & & & & $E_h/(e a_0)$ & & $5.14\times 10^{11} \ $V m$^{-1}$ \\
		\hline
		\end{tabular}
	\caption{Connection between the atomic units and the SI units.}
	\label{tab:AtomicUnits}
\end{table*}

\subsection{Grid choice}
\label{Sec:Grid}

The most obvious way to implement a grid method is to choose a spacing $\Delta x$ and a box size $L$.  The number of grid points is then $N\approx(L/\Delta x)^3$.  
Unfortunately this will be computationally prohibitive: if we want to consider a highly excited state of hydrogen $nl$, then the grid must go out to at least a 
radius of $2n^2$, and preferably much more, so $L > 4n^2$.  On the other hand, to resolve the $1s$ state properly a fine spacing (below $\sim 0.2$) is needed.  This leads us to the 
conclusion that we need $N\sim 4000n^6$ grid points, which is prohibitive for states above $n = 4$.

Clearly, we will need a type of grid that puts resolution where we need it: high resolution near the protons, and more modest resolution far away.  To avoid the 
complexity of developing an adaptive code, we will insist on high resolution near the trajectories of the proton and lower resolution elsewhere.  This immediately 
suggests developing a generalized cylindrical coordinate system, i.e., introducing a mapping $(u,v)\leftrightarrow(x,y)$ and using as our fundamental coordinates 
$(u,v,z)$ instead of $(x,y,z)$.  A constant grid spacing $\Delta u=\Delta v$ can then correspond to a variable spacing in the $xy$-plane, in accordance with the 
Jacobian of the transformation.  Note that high resolution (several grid points per Bohr radius) is also required in the region in between the protons in order to 
correctly model the $1s_{\rm A}\leftrightarrow1s_{\rm B}$ tunneling that is primarily responsible for charge transfer at low and intermediate velocities.

A minimal criterion for such a grid is that it should be able to adequately sample all bound wave functions with several grid points per cycle.  The momentum of a 
bound wave function can be as large as $\sqrt{2/\rho}$, where $\rho$ is the minimum separation from the nucleus.  Therefore the grid spacing should be at most 
$\sim\sqrt{\rho/2}$ (and preferably better). This requirement could be relaxed if $\rho\le 1$, where the classical intuition concerning the ``maximum momentum of a bound electron'' is invalid; in this regime instead all that is required is to have at least a few grid points per Bohr radius.


After considering and rejecting several other choices\footnote{For example, parabolic cylinder coordinates would have provided the desired resolution for head-on collisions, but at large impact parameter would have difficulty providing resolution at the locations of both protons.} we decided on the coordinate system
\begin{equation}
x = u\sqrt{1+\frac{u^2}{u_s^2}}, \;\;\;\; y=v\sqrt{1+\frac{v^2}4},
\end{equation}
where $u_s$ is a parameter.  The Jacobian is
\begin{equation}
\frac{\partial x}{\partial u} = \frac{1+2u^2/u_s^2}{\sqrt{1+u^2/u_s^2}} \rightarrow \left\{
\begin{array}{ll}
1 &  |u|,|x|\ll u_s \\
2u/u_s\approx \sqrt{8x/u_s} &  |u|,|x|\gg u_s
\end{array}\right.
\end{equation}
and
\begin{equation}
\frac{\partial y}{\partial v} = \frac{1+v^2/2}{\sqrt{1+v^2/4}} \rightarrow \left\{
\begin{array}{ll}
1 & |v|,|y|\ll 1 \\
v\approx \sqrt{2y} & |v|,|y|\gg 1.
\end{array}\right.
\end{equation}
This satisfies our resolution criteria if $\Delta u$ is no more than a few tenths, and $u_s$ is at least as large as $\sim\max(1,b)$. 

\begin{figure}
\includegraphics[width=3.2in]{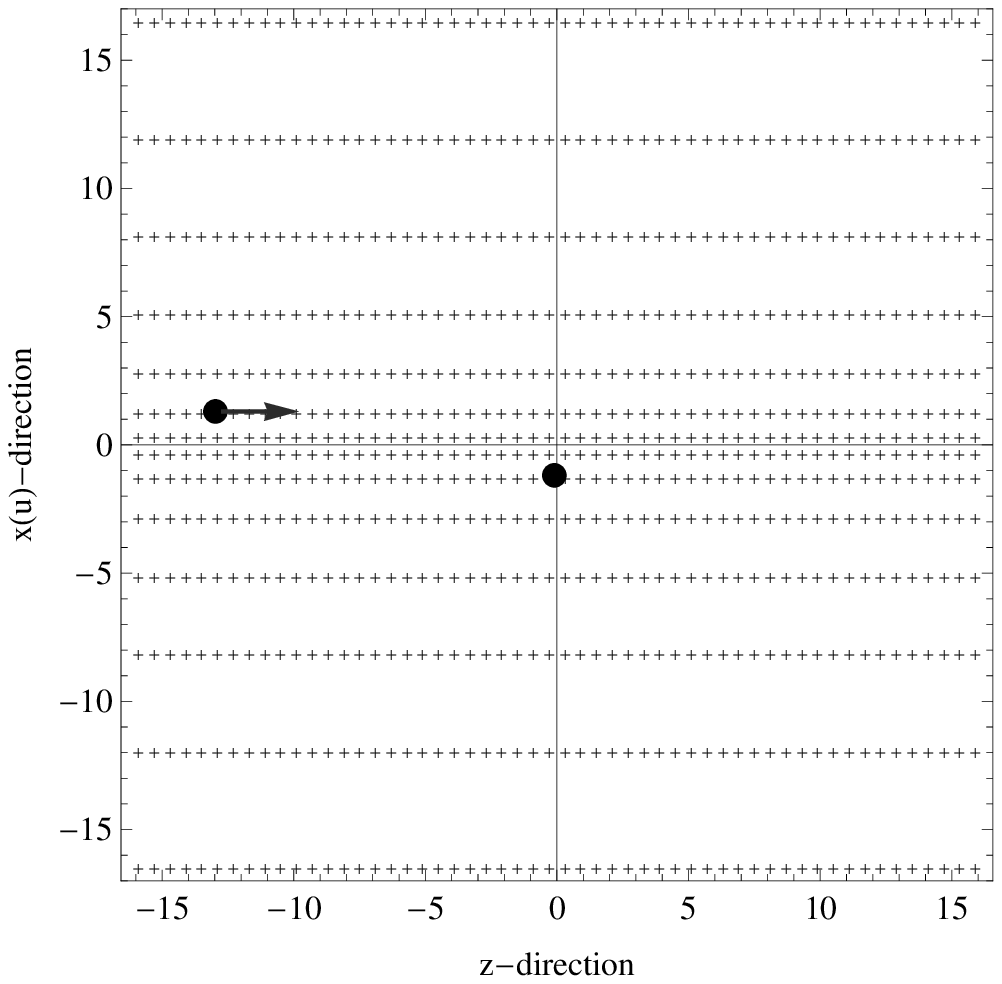}
\includegraphics[width=3.2in]{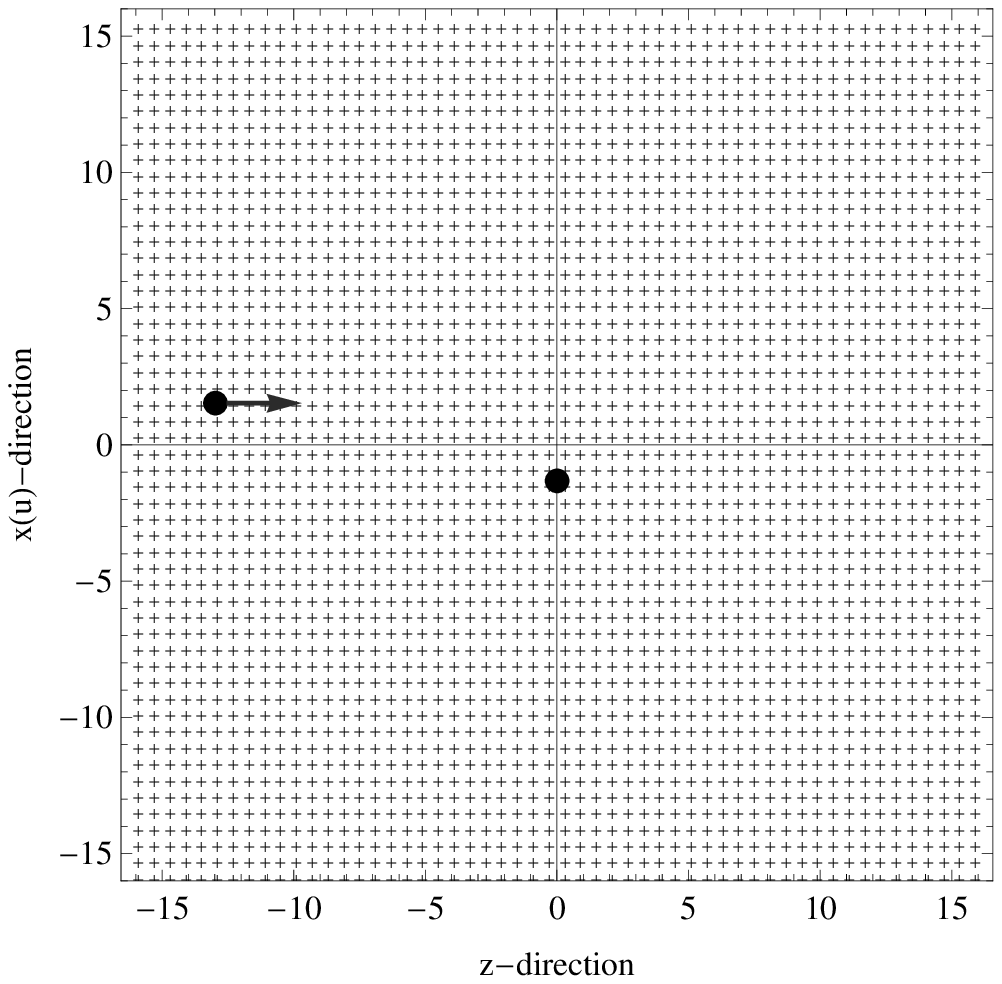}
\caption{\label{Fig:Grid} An example of a grid in $xz$-plane used in our analysis (upper panel) with every third point plotted in both $x$- and $z$-direction, so that the aspect ratio is illustrated correctly. We also show positions of the moving hydrogen atom (thick dot with the vector) and stationary proton at $z = 0$. Distances between grid points in $x$-direction correspond to equidistant intervals ($\Delta_u = 0.2$ in u-space), so that in the region close to the interacting particles $\Delta_x \approx \Delta_u = 0.2$, while at large separations spacing in $x$-direction is significantly increased reaching, for example, $\Delta_x \sim 1.6$ at $x \sim 16$. The lower panel shows a ``brute force'' Cartesian grid with $\Delta_x = \Delta_z = 0.2$ (again only every third point is plotted) and allows to clearly see the advantage of our grid choice.}
\end{figure}

We illustrate the prosed approach in Figure~\ref{Fig:Grid} by using a sample case of $\Delta_u = \Delta_z = 0.2$ and comparing our proposed grid to a ``brute force'' Cartesian grid with $\Delta_x = \Delta_z = 0.2$. This figure shows every third point in both $x$-direction and $z$-direction, and clearly illustrates benefits of putting high resolution in the region surrounding colliding particles while reducing the resolution far from the collision region. The resolution in the physical x-space between colliding particles is close to $\Delta_x \approx \Delta_u = 0.2$, while at larger separation the spacing between the grid points in $x$-direction significantly increases reaching $\Delta_x \sim 1.6$ at $x \sim 16$. 

By using a grid spacing of e.g. 0.2 in $(u,v,z)$ space, and noting that we only have to go out to maximum values of $u_{\rm max}\sim \sqrt{2u_s}n$, $v_{\rm 
max}\sim \sqrt2n$, the number of grid points necessary would be
\begin{equation}
N \sim\frac12(10\sqrt{2u_s}n)(10\sqrt2n)(20n^2) = 2000u_s^{1/2}n^4.
\end{equation}
This is much more manageable than $4000n^6$ as found earlier but is probably still too large for cases above $n \ge 4$.

We can make yet another improvement by Fourier-transforming the $z$-direction.  We suppose that we have a box of size $L_z$ in the $z$-direction and at each point in the $(u,v)$-plane we do a Fourier transform,
\begin{equation}
\Psi(x,y,z) = L_z^{-1} \sum_q \Psi_q(u,v) \rme^{2\pi\rmi q/L_z},
\end{equation}
where $q$ is an integer.  The maximum required value of $q$ is now $p_{z,\rm max}L_z/2\pi$.  Here $p_{z,\rm max}$ can be set to a large value ($p_{z,\rm max} \sim 20$) within a few Bohr radii of the atoms (i.e., $|u|,|v|$ less than a few), but a smaller value ($p_{z,\rm max} \sim 4$) at large separations. We refer to this as a q-drop procedure and it effectively reduces the number of grid points involved in the computation by removing Fourier modes that have very little contribution to the final result. For the results presented in this paper (cross sections with $n=4$ final principal quantum number), it is computationally feasible to do the calculations at a single $q_{\rm max}$. However the code we have written, \texttt{BDSCx}, supports the use of two values of $q_{\rm max}$ in different regions of the $(u,v)$-plane. The latter capability will be required for cross sections to the $n\ge 5$ levels of hydrogen. We are currently tuning the parameters of the code for this purpose, but this paper presents the $n\le 4$ results with a single $q_{\rm max}$ in order to make these available to the community in a more expedient way.



A peculiar property of this setup is that since we can only track momenta out to $p_{z,\rm max}$ it actually matters which nucleus we take as moving and which as fixed. One would expect that the best results would be obtained by taking the H atom as fixed and the H$^+$ ion as moving, but for charge transfer reactions the definitions of ``fixed'' and ``moving'' change. We prefer to handle charge transfer by applying a boost operation by an amount $V$ (i.e. increment the values of $q$ by $VL_z/2\pi$) to the wave function of the electron.  An alternative would be 
to evolve final states $|nlm_{\rm B}\rangle$ backward from $t=T\rightarrow 0$, apply a boost, and compute an inner product with the forward-evolved $|1s_{\rm A}\rangle$ state.  For Hermitian discretized Hamiltonians these two methods are equivalent.

To properly apply boost operation we consider two reference frames: first, a stationary frame K with our initial coordinates $(u,v,z)$ and the second is the restframe of a moving atom K$'$, which moves with the velocity $V$ in the $+z$ direction. The transformation is
\beq
\Psi(\vec{r},t) = \Psi'(\vec{r} - \vec{V}t,t) \rme^{\rmi(\vec{V}\cdot\vec{r} -V^2t/2)}.
\eeq

\subsection{Operators}

Here we restrict ourselves to orthogonal coordinate systems, i.e., where $\nabla u\cdot\nabla v=0$.

We next need a method to compute inner products, and Hermitian discretizations of the kinetic and potential operators.  The inner product is simply
\begin{equation}
\langle \varphi|\psi\rangle
= L_z^{-1} \sum_{quv} \varphi_q^\ast(u,v) \psi_q(u,v) \frac{\Delta u\,\Delta v}{\sqrt{g^{uu}g^{vv}}},
\label{eq:iprod}
\end{equation}
where $g^{uu} = |\nabla u|^2$, and $g^{vv} = |\nabla v|^2$.

The kinetic energy operator is the sum of the operators along the $x$, $y$, and $z$ axes.  The $z$-operator is trivial, being simply a 
multiplication by $2\pi^2q^2/L_z^2$.  The $x$ and $y$ operators are trickier; fortunately, they commute with the Fourier transform in the $z$ direction so we may 
implement them independently on each $q$-slice of the wave function.  We recall that the $x$ and $y$ components of the kinetic energy operator can be written as:
\begin{eqnarray}
\langle \varphi|\hat T_{xy}|\psi\rangle  =  \frac12
\int \left( \frac{\partial\varphi^\ast}{\partial x}\frac{\partial\psi}{\partial x}
  + \frac{\partial\varphi^\ast}{\partial y}\frac{\partial\psi}{\partial y} \right)\,\rmd^3{\mathbf r} =
\nonumber \\ 
\frac1{2L_z} \sum_q
\int \left( g^{uu}\frac{\partial\varphi^\ast}{\partial u}\frac{\partial\psi}{\partial u}
  + g^{vv}\frac{\partial\varphi^\ast}{\partial v}\frac{\partial\psi}{\partial v} \right)\frac{\rmd u\,\rmd v}{\sqrt{g^{uu}g^{vv}}}.
\end{eqnarray}
Thus, we see that the $u$ and $v$ parts of the kinetic energy operator are simply additive, i.e., we can write $\hat T=\hat T_u+\hat T_v+\hat T_z$.  The $u$-part can be 
re-cast by discretizing the partial derivative as
\begin{equation}
\frac{\partial\psi}{\partial u} = \frac{\psi(u+\Delta u/2) - \psi(u-\Delta u/2)}{\Delta u}.
\label{eq:pu}
\end{equation}
Note that this partial derivative is measured not on the grid points, but halfway in between (i.e., $\Delta u/2$ to the ``right'' of each grid point, or alternatively 
along each grid segment).  Then we may write $\langle \varphi|\hat T_u|\psi\rangle$ as a sum over such grid segments.  This gives an approximation to $\hat T_u$,
\begin{eqnarray}
\hat T_u\psi(u) \!\!\! &=& \!\!\! \frac{\sqrt{g^{uu}(u)g^{vv}(v)}}{2\Delta u^2} 
\nonumber \\ && \!\!\! \times
\Biggl\{
-\sqrt{\frac{g^{uu}\left(u+\frac{\Delta u}2\right)}{g^{vv}(v)}} [\psi(u+\Delta u)-\psi(u)]
\nonumber \\ && \!\!\!
+\sqrt{\frac{g^{uu}\left(u-\frac{\Delta u}2\right)}{g^{vv}(v)}} [\psi(u)-\psi(u-\Delta u)]
\Biggr\}.
\end{eqnarray}
A simple calculation shows that with the discretized inner product of Eq.~(\ref{eq:iprod}), this kinetic energy 
operator is exactly Hermitian.  A similar equation holds for $\hat T_v$.  Off-grid points are assumed to have $\psi=0$, corresponding to Dirichlet boundary conditions. We note, that versions of this operator with higher order accuracy can be constructed by using more than 2 points in the derivative described in Eq.~(\ref{eq:pu}). An intelligent boundary condition would have to be chosen at the endpoints; however, only the ionized electrons will reach the boundary and they will reflect off. The higher order derivatives can be used to reduce spurious oscillations between $ns \leftrightarrow np \leftrightarrow nd$ states. We relegate the study of the use of the higher order derivatives to a future work.   

The potential energy operator is local in 3-dimensional position space, but not in $(u,v,q)$-space.  The potential at any position is given by
\beqa
\nonumber
V(\mathbf r) = -\frac1{\sqrt{(x-b/2)^2+y^2+(z-z_{\rm A})^2}} - \\
\frac1{\sqrt{(x+b/2)^2+y^2+(z-z_{\rm B})^2}}.
\eeqa
The most efficient way to implement the potential operator is to FFT $\psi_q(u,v)$ in the $z$-direction, multiply by $V(x,y,z)$, and perform an inverse FFT.  For $N_z$ points, this implies $\sim ln N_z$ operations per grid point, which is manageable especially since in most cases $N_z$ is small because of the q-drop procedure.

We note that this choice of potential term in the Hamiltonian is associated with numerical difficulties due to its divergent nature near the proton. We eliminate these difficulties by capping the potential with a continuous function near the origin. We adopt
\begin{equation}
V(r) = \left\{\begin{array}{ll}
-\frac14R_0^{-1}(9-5R_0^{-2}r^2) & r<R_0 \\ -r^{-1} & r\ge R_0,
\end{array}\right.
\label{Eq:Potential}
\end{equation}
which was chosen so that the volumetric integral vanishes: $\int (V_{\rm capped} - V_{\rm true}) \rmd^3{\bmath r} = 0$. The advantage of this is that spurious features in the potential near the origin will result in a spurious interaction Hamiltonian between any two states $\psi_1$ and $\psi_2$ given by $\int \psi_1^\ast\psi_2 (V_{\rm capped} - V_{\rm true}) \rmd^3{\bmath r}$, and hence setting the volumetric integral to zero should yield improved behavior over, e.g., imposing a simple floor on $V$.
The capping radius is chosen to be $R_0 = 0.2$ for the results shown here. The capping procedure was tested extensively and shown not to introduce any spurious deviations from the results with uncapped potential. Details of the testing procedures are discussed in Subsection~\ref{Sec:Tests} and illustrated in Fig.~\ref{Fig:Check}. For all relevant quantities calculated using our code the difference between capped and uncapped potential is $<< 1\%$.

\section{The code}
\label{Sec:Code}

\subsection{Grid parameters}

To implement the proposed algorithm we developed a grid-based code \texttt{BDSCx}, which computes cross sections of hydrogen-proton collisions with hydrogen starting in $|1s\rangle$ state:
\beq
|1s\rangle = \Psi_{100} = \frac{1}{\sqrt\pi}e^{-r}.
\eeq

\begin{figure*}
\centering
\includegraphics[width=3.4in]{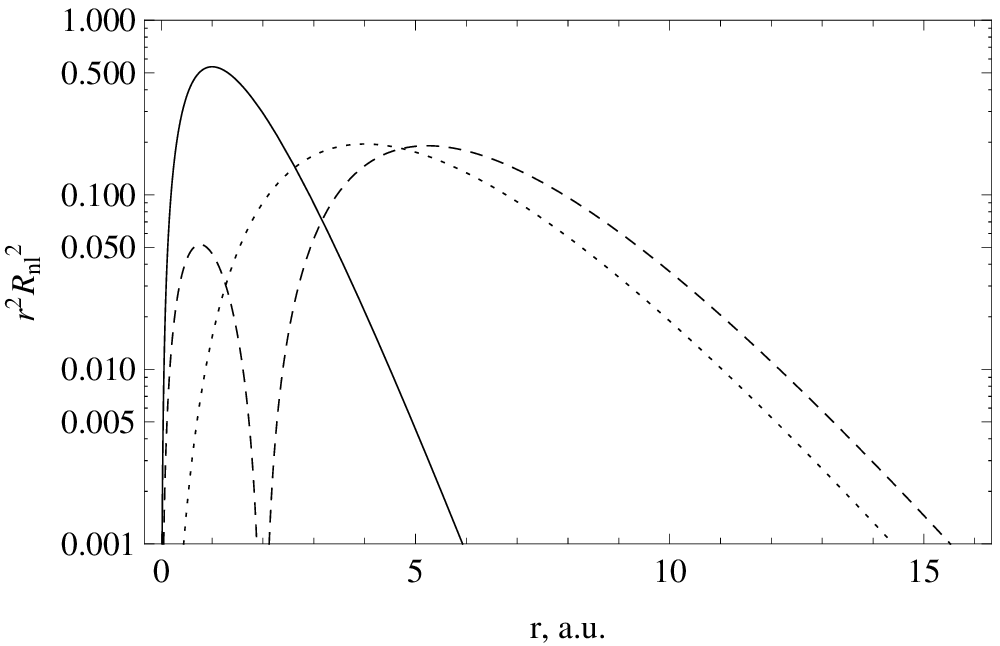}
\includegraphics[width=3.4in]{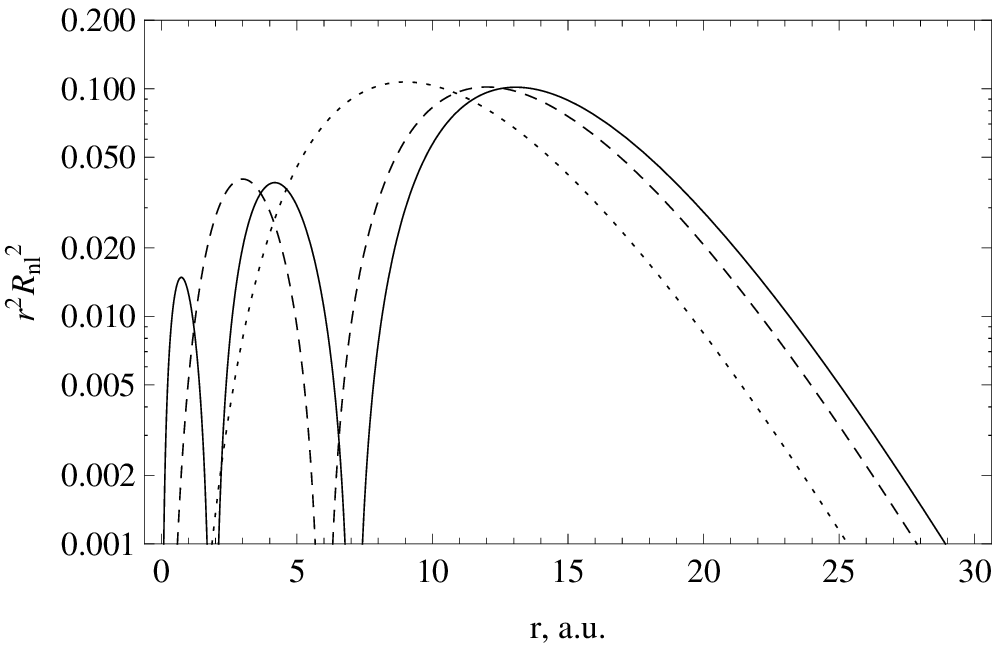}
\caption{\label{Fig:Eigen} Plot on the left shows charge density distribution $r^2 R_{nl}^2(r)$ for $1s$ (solid), $2s$ (dashed), and $2p$ (dotted) states. Plot on the right shows charge distribution for $3s$ (solid), $3p$ (dashed) and $3d$ (dotted) states.}
\end{figure*}

We start and end the collision when the the two particles A and B are separated by a distance sufficient to fully resolve wave functions of interest. Distance from the edges of the box is also determined by the requirement to properly resolve $nlm$ states of interest at the beginning and at the end of the collision. Our choice of the box size parameters is guided by the charge distribution in the states of interest. In Figure~\ref{Fig:Eigen}, we plot the charge density $r^2 R_{nl}^2(r)$ for hydrogen states with $n=1,2,3$. Here, $R_{nl}(r)$ are normalized radial eigenfunctions of the hydrogen atom:
\begin{equation}
R_{nl}(r) = \sqrt{ \frac{(n - l - 1)!}{4n^4[(n+l)!]^3}}
\,\rme^{-r/n}\left(\frac{2r}n\right)^l L^{2l+1}_{n-l-1}\left(\frac{2r}n\right),
\label{Eq:Rnl}
\end{equation}
where $L^{2l+1}_{n-l-1}(2r/n)$ are the generalized Laguerre polynomials.

We choose the length of the box in the $z$-direction by requiring charge density decrease of more than 2 orders of magnitude relative to its maximum value for the state of interest. For example, to properly resolve the $1s$ state a particle should be separated from the edge of the box by more than 5 Bohr radii, whereas for the $n=2$ we need more than 15 Bohr radii of separation. These conservative resolution requirements lead to the size of the box in z direction given by $L_z = 5 + 15\times 4 = 65$ in the case when only $n=2$ states are of interest.  

We also use the same conservative requirements for the box size in the $x$ ($u$) and $y$ ($v$) directions. The size in the $y$ direction is fixed as soon as we decide on the upper $n$ state of interest, while the size in $x$ direction also depends on the impact parameter $b$ so that $L_x = L_y + b$. We further require high resolution near the particles so  that $\Delta u \sim \Delta v \sim \Delta z \sim 0.18$. This resolution requirement was tested and found to converge with the difference between $\Delta \sim 0.18$ and $\Delta \sim 0.16$ being less than $0.1$ per cent. Several examples of input parameters required for accurate cross section results are provided in Table~\ref{tab:SimParam}.

\begin{table*}
	\centering
		\begin{tabular}{ | l | c | c | c | c | c | c | c | r | }
		\hline
 $n$ & $b$ & $u_s$ & $L_x (L_u)$ & $L_y (L_v)$ & $L_z$ & $N_u$ & $N_v$ & $N_z$ \\
 \hline
 $2$ & $ 1$ & $ 1 $ & $31 (8)$ & $30 (11)$ & $65$ & $46$ & $62$ & $362$ \\ 
 \hline 	 	
 $2$ & $ 5 $ & $ 5 $ & $35 (18)$ & $30 (11)$ & $65$ & $100$ & $62$ & $362$ \\ 
 \hline 	 	
 $4$ & $ 1$ & $ 1 $ & $101 (15)$ & $100 (20)$ & $205$ & $84$ & $112$ & $1140$ \\ 
 \hline 	 	
 $4$ & $ 5 $ & $ 5 $ & $105 (32)$ & $100 (20)$ & $205$ & $178$ & $112$ & $1140$ \\ 
 \hline
 $5$ & $ 1$ & $ 1 $ & $143 (18)$ & $142 (24)$ & $285$ & $100$ & $134$ & $1584$ \\ 
 \hline 	 	
 $5$ & $ 5 $ & $ 5 $ & $147 (38)$ & $142 (24)$ & $285$ & $212$ & $134$ & $1584$ \\ 
 \hline 	 	
	\end{tabular}
	\caption{Examples of the simulation parameters for collisions involving $n = 2$ and $n=4$ states.}
	\label{tab:SimParam}
\end{table*}

\begin{figure*}
\centering
\includegraphics[width=2in]{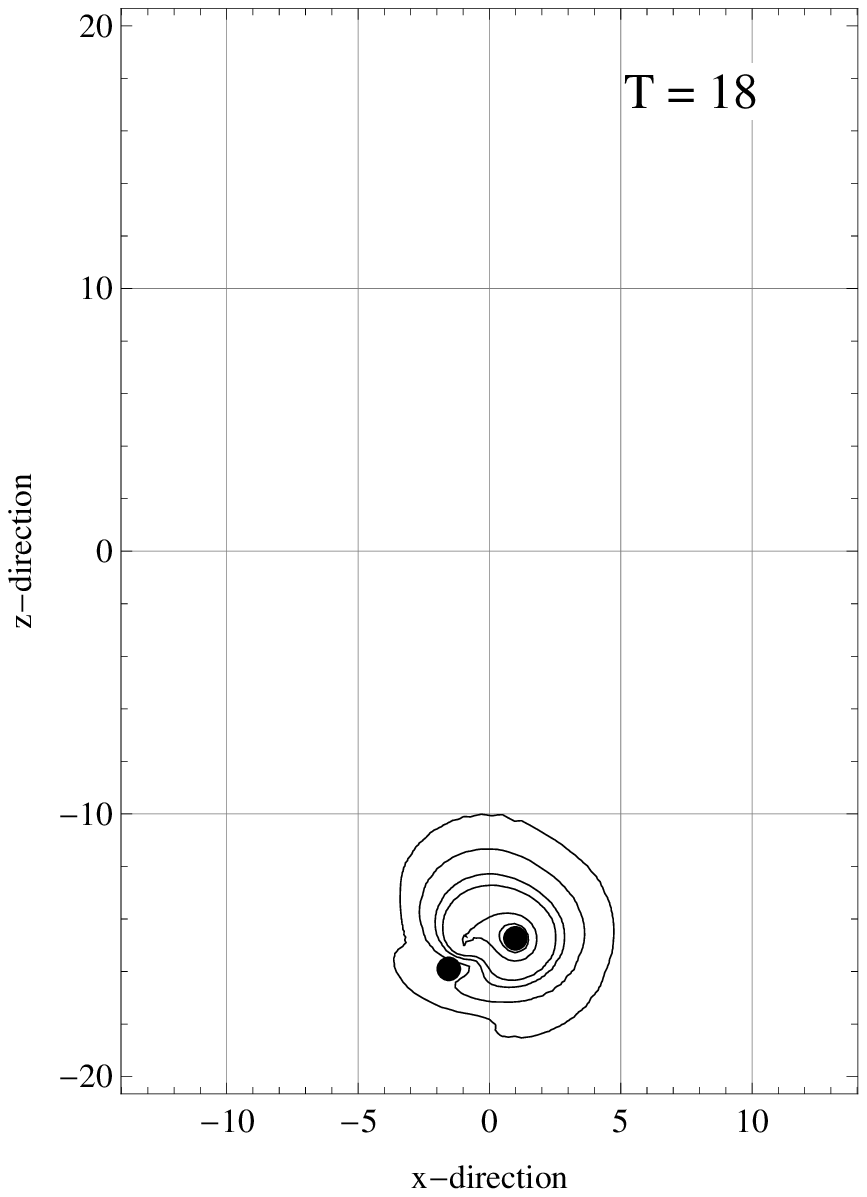}
\includegraphics[width=2in]{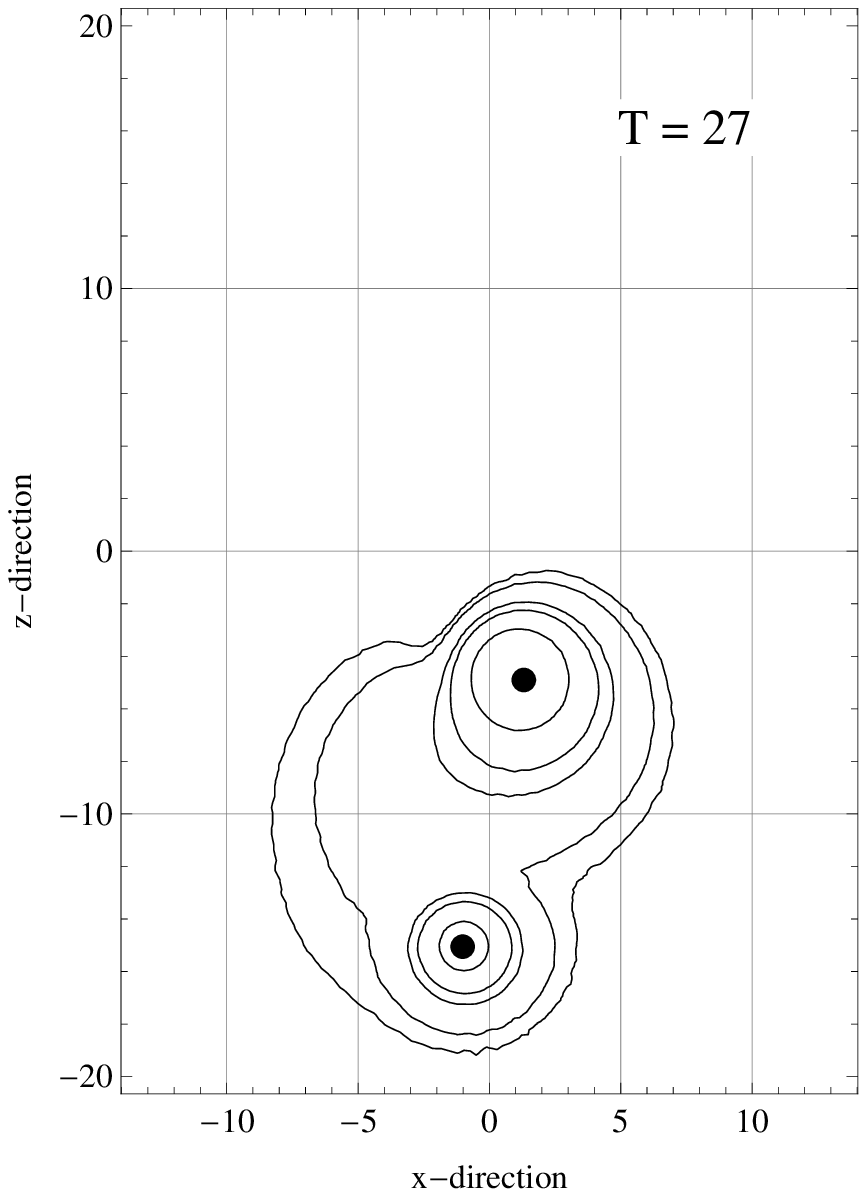}
\includegraphics[width=2in]{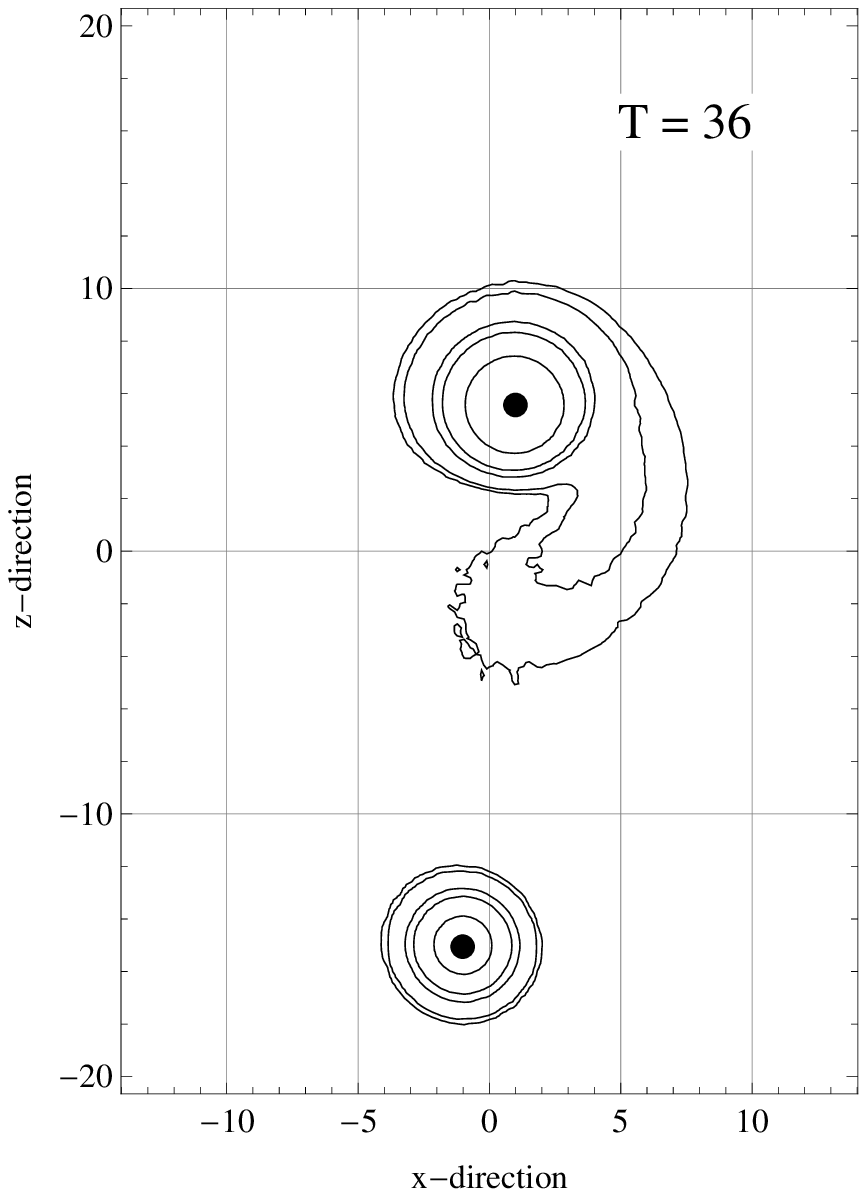}
\caption{\label{Fig:Psi}Contour plots showing time evolution of the electron probability density as hydrogen atom moves with $V = 1$ and $b=2$ past the proton at rest. Time in the panels is given in atomic units.}
\end{figure*}

One of the important advantages of the grid approach chosen in our calculations is the ability to visualize the evolution of the wave functions during the collision. In Figure~\ref{Fig:Psi}, we show the time evolution of electron's probability density as the hydrogen atom moves past the stationary proton with impact parameter $b = 2$ and velovity $V = 1$ a.u. We clearly see that after the collision part of electron's probability density is spread between the two atoms, indicating possible charge transfer during the impact. 

\subsection{Collisions with large impact parameter}

At large values of the impact parameter $b$ the number of required grid points in the $u$ and $v$ directions increases dramatically. Fortunately, at large separations ($ b \geq 5$) the cross section results are well described by the Born approximation if the collision energies are reasonably high ($E_{\rm col} \geq 1$ keV). To reduce the amount of numerical computations without sacrificing precision of the cross-section results we use the Born approximation to obtain transition probabilities above $b = 5$. It is easy to show that in this approximation the transition probability into the $X$ final state can be written as:
\begin{equation}
|\langle X|\hat S(-T,T)|1s\rangle |^2 =
\left|\int_{-\infty}^{\infty} \rme^{-\rmi(E_X - E_0)t}\langle X|\hat{W}|1s\rangle \,\rmd t\right|^2,
\end{equation}
where $\hat{W}$ is the perturbation Hamiltonian operator, which in our case is given by
\beq
\hat{W} = \frac1R-\frac{1}{|\vec{R} - \vec{r}|},
\eeq
and $\vec{R}$ is the vector separating the two protons. If one approximates the Hamiltonian as a dipole,
$\hat{W} \approx -\vec r\cdot\vec R/|R|^3$,
as appropriate at large impact parameters, then an analytic solution for the transition probability is obtained. In the first order, probability of excitation into $|X>$ can be written as:
\beqa
|\langle X|\hat S(-\infty,\infty)|1s\rangle|^2 \approx |-i \int^{\infty}_{-\infty} e^{-i(E_f - E_0)\xi/v}\times \\ \nonumber
(\frac{b}{(b^2 + \xi^2)^{3/2}}\langle X|x|1s\rangle + \frac{\xi}{(b^2 + \xi^2)^{3/2}}\langle X|z|1s\rangle)\frac{d\xi}{v}|^2.
\eeqa

For example, the first order probability of transition into $2p_0$ state becomes
\beq
|\langle 2p_0|\hat S(-\infty,\infty)|1s\rangle|^2 \approx \left(\frac{32\sqrt{2}K_0(\frac{3 b}{8 v})}{81 v^2}\right)^2,
\eeq
where $K_0(r)$ is a Bessel function of the second kind.  

Note that the Born approximation only allows transitions into the $|np_{\rm A}\rangle$ final states (and to ionized final states).

At collision energies of $E_{\rm col} > 1$ keV and impact parameters of $b > 5$, the results produced by using the Born approximation differ from the results obtained by our code by less than $5 $ per cent, and because of the small value of the transition probabilities at large $b$ values the difference introduced into cross section results is $\ll 1$ per cent.

\subsection{Convergence and consistency tests}
\label{Sec:Tests}

\begin{figure}
\includegraphics[width=3.4in]{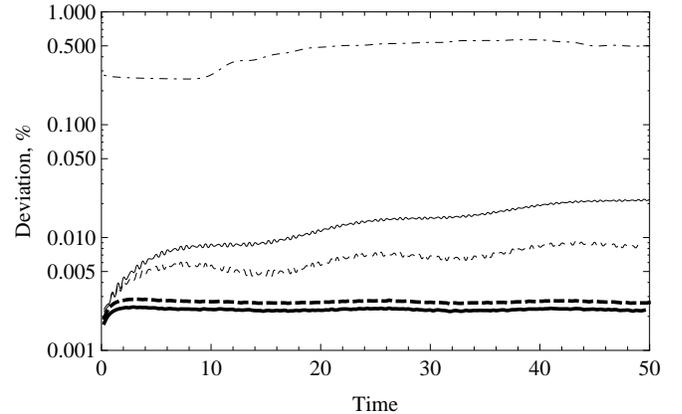}
\caption{\label{Fig:Check} Deviation of $|\langle 1s | \hat S(T) | 1s \rangle|$ from unity as a function of time in a test runs without the second particle. Thick curves show the evolution of a stationary hydrogen atom located at $x = 0.5$, $y = 0$ and $z = -20$ for the cases with unmodified $1/r$ potential (solid) and modified potential of Eq.~\ref{Eq:Potential} (dashed). Regular curves show the evolution of a hydrogen moving in the positive $z$-direction with $V = 1$ with unmodified potential (solid) and capped potential (dashed). Dash-dotted curve shows deviation from the exact energy expectation $\langle H_0\rangle = -1/(2n^2)$ for stationary hydrogen.}
\end{figure}

To test the proper functionality of our code and fully understand the dependence of the results on the input parameters such as grid resolution, the size of the box within which we keep a large number of point in q-space, box size and others we ran a large number of tests most important of which are: 
\begin{enumerate}
	\item Time-dependent evolution of an electron in the field of a single proton at rest. In this case we would expect that the state of a system remains unchanged, i.e., if a system starts in $| 1s \rangle$ state it remains there and the probabilities of transitions into states with $n = 2$ vanish within the numerical precision of a discretized Hamiltonian system. Figure~\ref{Fig:Check} shows the result of system's evolution over a few dynamical times ($T_{dyn} = 2\pi n^3$). We clearly see that the system exhibits small oscillations associated with discretization of the Hamiltonian; however, the results remain close to the expected values for $|\langle 1s | S(T) | 1s \rangle|^2$ and for the energy expectation $\langle H_0\rangle$. For example, $|\langle 1s | \hat S(T) | 1s \rangle|^2 \approx 1$ over the full time period with the precision better than 0.05 per cent; energy expectation deviates from the theoretical prediction $\langle H_0\rangle = -\frac12n^{-2}$ by less than $0.5\%$. The probability of $|1s\rangle$ to $|2s\rangle$ transition remains small and oscillates with the period $T_{\rm osc} \sim 3\pi/4$ determined by the energy splitting between $n = 1$ and $n = 2$ levels. Furthermore, the precision of the results increases as we increase the resolution of the grid, allowing us to achieve desired accuracy of the final results through convergence.
	\item The next test incorporates the motion of the hydrogen atom. In this case, the initial wave function of the electron is phase-shifted by the factor $\rme^{\rmi Vz}$ as discussed at the end of Sec.~\ref{Sec:Grid}. The results of these tests are also shown in Figure~\ref{Fig:Check}. We clearly see that the normalization of the $|1s\rangle$ is properly conserved. We note, however, that the motion of a particle introduces small additional oscillations into the shapes of the curves.  
  \item We tested convergence of our cross section results by running our code with three different resolution values: $\Delta = 0.175, 0.18$ and $0.22$ at collision energy $E_{col} = 80$ keV. The results converge, with cross section difference between the $\Delta = 0.175$ and $\Delta = 0.18$ cases being less than 3\%. We also checked that increasing the box size at a fixed resolution does not modify our results, which indicates that the results of the boundary interactions are minimal for the set of parameters used in our runs. 
\end{enumerate}

It is important to note that, while our cross section results converge within the limits of simulation accuracy, they still contain inherent uncertainty associated with our discretization procedure, finite resolution of the numerical grid, finite size of the collision box, approximation of straight line trajectory for colliding protons, reconstruction of the cross sections from discrete probability values on a grid of impact parameters $b$, and the use of the Born approximation at high values of impact parameters. Numerical errors associated with these sources are extremely hard to quantify precisely, although for each individual source the error has been tested and minimized with the convergence procedure to be below 1\%. 

In testing we also compared our predictions for low $n$ ($n \leq 3$) cross sections with the results of previous studies and showed that our results are fully consistent with the results presented in~\cite{Kuang96},~\cite{Kolakowska99}, and~\cite{Winter09}. We note, however, that the major goal of this paper is to introduce the computational algorithm and the code that allow accurate cross section calculations at high values of $n$ as well as illustrate the importance of these calculations in studying Balmer-dominated shocks. Therefore, detailed comparison of our results to the results of other studies is not performed here and will be presented in future papers.

\section{Cross section results}
\label{Sec:Results}

In this section, we present results for the cross sections for excitation and charge transfer final states, and (where possible) compare our results to previous computations.

We obtain cross sections by integrating Eq.~(\ref{eq:sigma-X}), with $|X\rangle = |nlm\rangle$. The final state wave functions in the rest frame of the final atom (A or B) are given by
\begin{equation}
	|nlm\rangle = \Psi_{nlm}(r,\theta,\phi) = R_{nl}(r) Y^m_l(\theta,\phi).
\end{equation}
Here, $R_{nl}(r)$ are normalized radial eigenfunctions (Eq.~\ref{Eq:Rnl}) and $Y^m_l(\theta,\phi)$ are spherical harmonics. If the final atom is moving, then a boost must be applied to the relevant wave function.

\begin{figure*}
\centering
\includegraphics[width=3.4in]{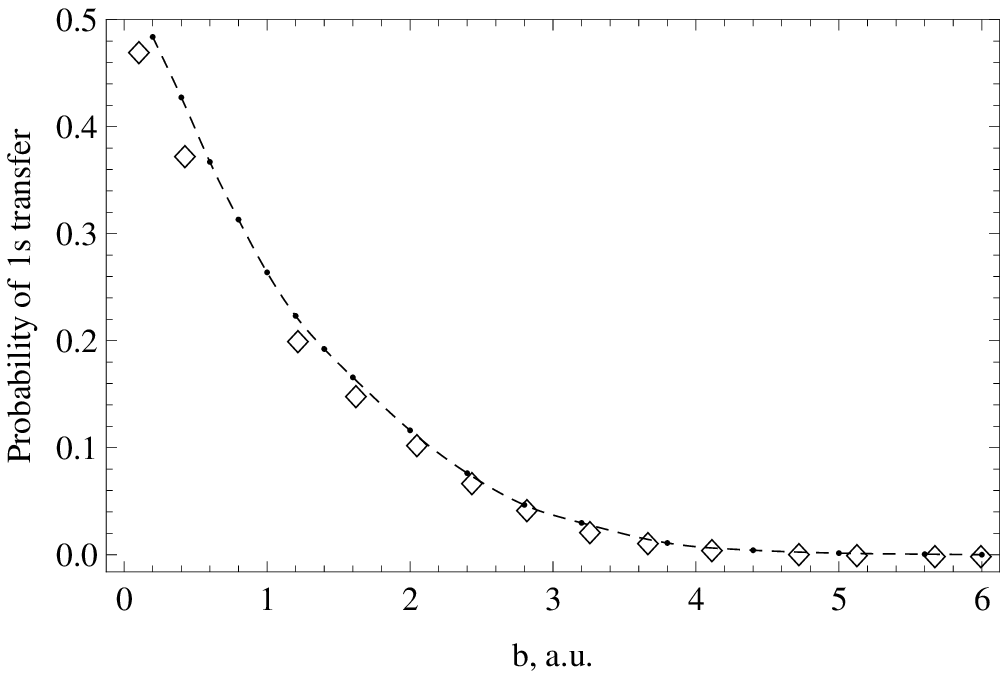}
\includegraphics[width=3.4in]{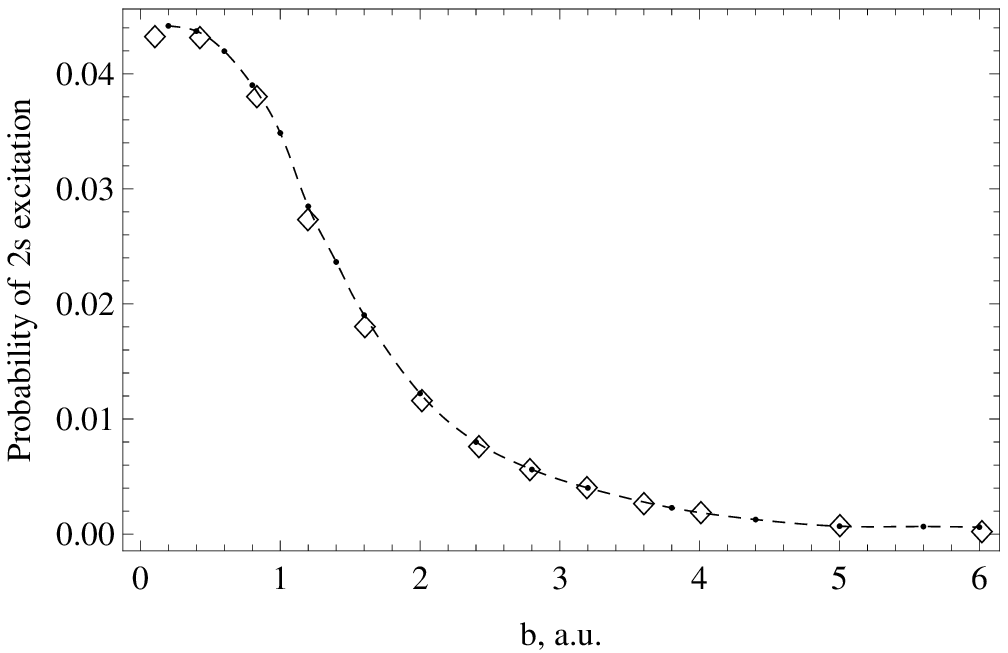}
\caption{\label{Fig:Prob} Plot on the left shows probability of charge transfer into 1s state as a function of impact parameter for $E_{col} = 40$ keV, and the plot on the right shows probability of excitation into 2s state. Diamonds show corresponding results from \citet{Kolakowska98}.}
\end{figure*}

The first results obtained by running our code correspond to collisions involving only $n = 1$ and $n = 2$ states. This problem is the least computationally demanding and it has a large amount of data produced by previous studies, allowing for further tests of the code. In Figure~\ref{Fig:Prob} we plot probabilities of excitation to $2s$ state and charge transfer into $1s$ state in collisions with impact energy $E_{\rm col} = 40$ keV. We compare our results with the results obtained in~\cite{Kolakowska98} and show good agreement over the full range of $b$ with slight deviation at small values of impact parameter caused by higher resolution used in our runs. Our deviation in the case of charge transfer into $1s$ state causes higher value of the overall cross section, which is consistent with the results obtained by other groups (see for example~\cite{Winter09} and references therein).

\begin{figure}
\centering
\includegraphics[width=3.4in]{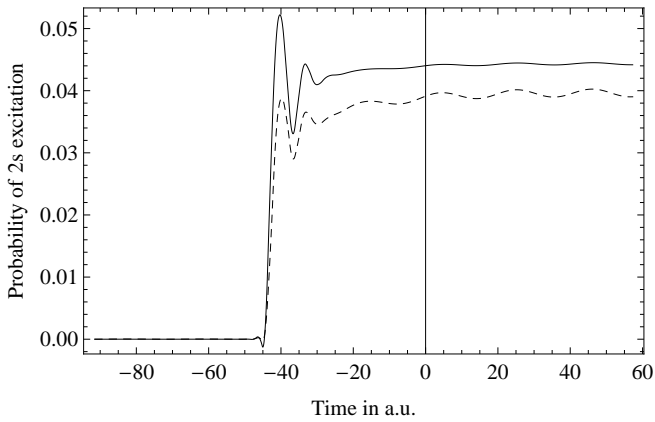}
\caption{\label{Fig:Prob2s} Probability of excitations into $2s$ states as a function of time for 2 different values of impact parameter $b = 0.2$ a.u. (solid curve) and $b = 0.8$ a.u. (dashed curve).}
\end{figure}

A significant advantage of using direct solution of the Schr\"odinger equation on a grid is the ability to see the convergence of the probability results. In Figure~\ref{Fig:Prob2s} we show an example of probability evolution for excitations into $2s$ states. This method clearly allows to see the convergence of the numerical calculation and allows direct comparison of runs with various box parameters. 

\begin{figure}
\centering
\includegraphics[width=3.4in]{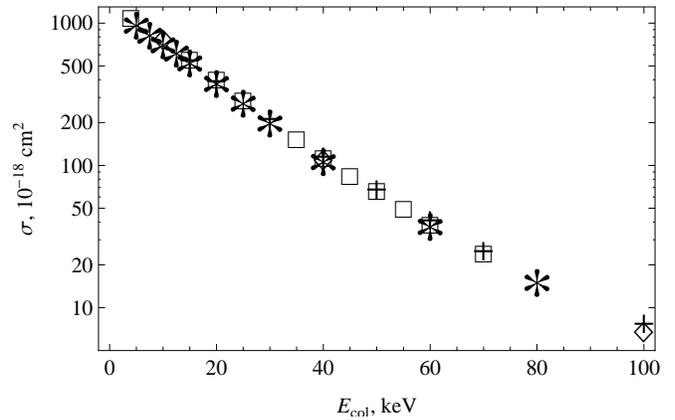}
\caption{\label{Fig:Sign2} Excitation cross sections (in units of $10^{-18}$ cm$^2$) for charge transfer into $1s$ state. Stars show results of our calculations, diamonds correspond to Kolakowska (1998), squares show results from Winter (2009) and plus signs correspond to the results from Kuang \& Lin (1996).}
\end{figure}

We further note that our cross section results are fully consistent with the results obtained by other groups for low $n$ cases. For example, in Figure~\ref{Fig:Sign2} we plot cross section for charge transfer into $1s$ state for various collision energies and compare our results with other theoretical calculations \citep{Kuang96, Kolakowska98, Winter09}.

In this study we have limited our cross section calculation to $n = 4$, which allows us to calculate Balmer decrement discussed in the next section. Our results, showing the cross sections of charge transfer and excitations are provided in Tables~\ref{Tab:Xs} and~\ref{Tab:XsTrans}. In Appendixes A and B, we also provide Chebyshev polynomial fits to our results for excitations and charge transfer into $3s$, $3p$, $3d$, $4s$, $4p$, $4d$, and $4f$ states and plots of obtained cross sections compared to the results from~\cite{Kolakowska98} and~\cite{Winter09}. We note, that the \texttt{BDSCx} code produces the cross-sections for transitions into states with various $n$, $l$, and $m$, allowing for studies of polarization-dependent signals. Calculation of higher $n$ states, as well as more detailed analysis of the $q$-drop procedure are relegated to the future work. 

\begin{table*}
	\centering
		\begin{tabular}{ | l | c | c | c | c | c | c | c | c | c | c | c | r | }
		\hline
 $\sigma/V$ & $2s$ & $2p_0$ & $2p_{\pm 1}$ & $3s$ & $3p_0$ & $3p_{\pm 1}$ & $3d_0$ & $3d_{\pm 1}$ & $3d_{\pm 2}$ & \\
 \hline
 $5 \ keV$ & $6$ & $3$ & $13$ & $0.35$ & $0.4$ & $0.65$ & $0.18$ & $1.3$ & $0.01$ & \\
 \hline
 $7.5 \ keV$ & $4.6$ & $4.5$ & $9.1$ & $0.7$ & $0.45$ & $1.1$ & $0.29$ & $0.88$ & $0.004$ & \\
 \hline 
 $10 \ keV$ & $4.1$ & $5.7$ & $6.9$ & $0.8$ & $0.6$ & $1.3$ & $0.3$ & $0.6$ & $0.009$ &\\
 \hline
 $12.5 \ keV$ & $6.0$ & $7.8$ & $6.2$ & $1.0$ & $1.1$ & $1.1$ & $0.52$ & $0.46$ & $0.02$ &\\
 \hline 	 	
 $15 \ keV$ & 8.2 & 10.1 & 7.5 & 1.4 & 1.5 & 1.2 & 0.56 & 0.49 & 0.04 & \\
 \hline
 $20 \ keV$ & $12.0$ & $15.1$ & $10.4$ & $2.2$ & $1.9$ & $1.8$ & $0.8$ & $0.6$ & $0.07$ &\\
 \hline
 $25 \ keV$ & 15.9 & 19.1 & 14.2 & 3.3 & 2.7 & 2.2 & 1.1 & 0.69 & 0.11 &\\
 \hline
 $30 \ keV$ &  16.6 & 23.5 & 16.9 & 3.5 & 3.3 & 2.6 & 1.15 & 0.70 & 0.14 &\\
 \hline  	 	
 $40 \ keV$ & $15.8$ & $26.5$ & $19.5$ & $3.7$ & $4.3$ & $3.3$ & $1.1$ & $0.7$ & $0.15$ &\\
 \hline
 $60 \ keV$ & 13.9 & 28.3 & 23.5 & 3.0 & 4.4 & 3.9 & 0.9 & 0.5 & 0.19 & \\
 \hline 	 	
 $80 \ keV$ & $11.5$ & $27$ & $23$ & $2.7$ & $4.4$ & $4.1$ & $0.6$ & $0.4$ & $0.18$ &\\
 \hline 	 	
 \hline
  & $4s$ & $4p_0$ & $4p_{\pm 1}$ & $4d_0$ & $4d_{\pm 1}$ & $4d_{\pm 2}$ & $4f_0$ & $4f_{\pm 1}$ & $4f_{\pm 2}$ & $4f_{\pm 3}$ \\
 \hline
 $5 \ keV$ & $0.1$ & $0.15$ & $0.08$ & $0.1$ & $0.35$ & $0.002$ & $0.03$ & $0.2$ & $0.004$ & $2\times10^{-4}$ \\
 \hline
 $7.5 \ keV$ & $0.30$ & $0.14$ & $0.20$ & $0.11$ & $0.41$ & $0.001$ & $0.035$ & $0.13$ & $9\times10^{-4}$ & $1\times10^{-5}$ \\
 \hline
 $10 \ keV$ & $0.35$ & $0.17$ & $0.36$ & $0.16$ & $0.34$ & $0.003$ & $0.04$ & $0.08$ & $8\times10^{-4}$ & $4\times10^{-5}$ \\
 \hline
 $12.5 \ keV$ & $0.40$ & $0.37$ & $0.33$ & $0.21$ & $0.26$ & $0.007$ & $0.05$ & $0.05$ & $0.002$ & $1\times10^{-4}$ \\
 \hline
 $15 \ keV$ & 0.45 & 0.50 & 0.40 & 0.33 & 0.25 & 0.02 & 0.06 & 0.04 & 0.004 & $2\times10^{-4}$ \\
 \hline
 $20 \ keV$ & $0.8$ & $0.75$ & $0.58$ & $0.35$ & $0.29$ & $0.03$ & $0.07$ & $0.04$ & $0.005$ & $8\times10^{-4}$ \\
 \hline
 $25 \ keV$ &  1.1 & 1.0 & 0.69 & 0.54 & 0.35 & 0.05 & 0.08 & 0.05 & 0.007 & $6\times10^{-4}$ \\
 \hline
 $30 \ keV$ &  1.2 & 1.15 & 0.88 & 0.59 & 0.36 & 0.06 & 0.08 & 0.05 & 0.007 & $8\times10^{-4}$ \\ 
 \hline
 $40 \ keV$ & $1.35$ & $1.5$ & $1.1$ & $0.6$ & $0.4$ & $0.07$ & $0.06$ & $0.04$ & $0.007$ & $8\times10^{-4}$ \\
 \hline
 $60 \ keV$ & 1.15 & 1.6 & 1.4 & 0.5 & 0.3 & 0.09 & 0.03 & 0.02 & 0.005 & 0.001 \\
 \hline
 $80 \ keV$ & $1.1$ & $1.6$ & $1.5$ & $0.3$ & $0.2$ & $0.08$ & $0.015$ & $0.008$ & $0.003$ & $0.001$ \\
 \hline
		\end{tabular}
	\caption{Cross section results for excitation transitions into $n = 2,\ 3$ levels.}
	\label{Tab:Xs}
\end{table*}

\begin{table*}
	\centering
		\begin{tabular}{ | l | c | c | c | c | c | c | c | c | c | c | c | r | }
		\hline
 $\sigma/V$ & $1s$ & $2s$ & $2p_0$ & $2p_{\pm 1}$ & $3s$ & $3p_0$ & $3p_{\pm 1}$ & $3d_0$ & $3d_{\pm 1}$ & $3d_{\pm 2}$ \\
 \hline
 $5 \ keV$ & $1092$ & $5.8$ & $2.1$ & $11.5$ & $0.25$ & $0.35$ & $0.60$ & $0.20$ & $1.0$ & $0.008$ \\
 \hline 	 	 
 $7.5 \ keV$ & $928$ & $12$ & $3.0$ & $12.9$ & $0.55$ & $0.67$ & $0.89$ & $0.25$ & $1.2$ & $0.03$ \\
 \hline
 $10 \ keV$ & $795$ & $18$ & $4.9$ & $13$ & $1.5$ & $0.90$ & $1.5$ & $0.30$ & $1.1$ & $0.05$\\
 \hline
 $12.5 \ keV$ & $695$ & $27.1$ & $6.6$ & $11.6$ & $3.2$ & $1.6$ & $1.6$ & $0.40$ & $0.80$ & $0.04$\\
 \hline 	 	 	 	
 $15 \ keV$ & 593 & 32.8 & 7.8 & 9.8 & 4.8 & 2.1 & 1.6 & 0.38 & 0.52 & 0.03 \\
 \hline 	 	
 $20 \ keV$ & $425$ & $39$ & $7.9$ & $6.5$ & $8.6$ & $2.5$ & $1.7$ & $0.35$ & $0.28$ & $0.03$\\
 \hline	
 $25 \ keV$ & 309 & 38.6 & 6.8 & 4.4 & 8.9 & 2.4 & 1.0 & 0.28 & 0.12 & 0.015 \\
 \hline 	 
 $30 \ keV$ & 224 & 34.9 & 5.7 & 3.0 & 8.8 & 2.0 & 0.80 & 0.21 & 0.07 & 0.01 \\
 \hline 		 	
 $40 \ keV$ & $120$ & $22$ & $3.6$ & $1.5$ & $6.5$ & $1.3$ & $0.40$ & $0.10$ & $0.03$ & $0.005$\\
 \hline
 $60 \ keV$ & 42 & 8.6 & 1.3 & 0.40 & 2.7 & 0.50 & 0.15 & 0.03 & 0.01 & 0.002 \\
 \hline 	 	
 $80 \ keV$ & $17$ & $3.5$ & $0.49$ & $0.15$ & $1.1$ & $0.20$ & $0.05$ & $0.01$ & $0.004$ & $5\times10^{-4}$\\
 \hline 	 	
 \hline
  & $4s$ & $4p_0$ & $4p_{\pm 1}$ & $4d_0$ & $4d_{\pm 1}$ & $4d_{\pm 2}$ & $4f_0$ & $4f_{\pm 1}$ & $4f_{\pm 2}$ & $4f_{\pm 3}$ \\
 \hline
 $5 \ keV$ & $0.12$ & $0.20$ & $0.10$ & $0.20$ & $0.30$ & $0.003$ & $0.05$ & $0.20$ & $0.001$ & $2\times10^{-5}$ \\
 \hline
 $7.5 \ keV$ & $0.086$ & $0.29$ & $0.14$ & $0.07$ & $0.40$ & $0.008$ & $0.04$ & $0.17$ & $0.005$ & $8\times10^{-5}$ \\
 \hline
 $10 \ keV$ & $0.64$ & $0.39$ & $0.34$ & $0.11$ & $0.50$ & $0.02$ & $0.04$ & $0.11$ & $0.009$ & $1\times10^{-4}$ \\
 \hline
 $12.5 \ keV$ & $0.86$ & $0.70$ & $0.45$ & $0.22$ & $0.40$ & $0.016$ & $0.039$ & $0.08$ & $0.005$ & $1\times10^{-4}$\\
 \hline 	 	 	 	
 $15 \ keV$ & 1.46 & 0.94 & 0.51 & 0.29 & 0.30 & 0.01 & 0.035 & 0.05 & 0.003 & $1\times10^{-4}$ \\
 \hline
 $20 \ keV$ & $2.9$ & $1.2$ & $0.60$ & $0.30$ & $0.17$ & $0.02$ & $0.03$ & $0.02$ & $0.003$ & $6\times10^{-4}$ \\
 \hline
 $25 \ keV$ &  3.5 & 1.2 & 0.38 & 0.24 & 0.08 & 0.008 & 0.02 & 0.005 & $6\times10^{-4}$ & $5\times10^{-5}$ \\
 \hline	
 $30 \ keV$ &  3.6 & 1.0 & 0.30 & 0.20 & 0.05 & 0.005 & 0.01 & 0.002 & $3\times10^{-4}$ & $6\times10^{-5}$ \\
 \hline
 $40 \ keV$ & $2.7$ & $0.65$ & $0.17$ & $0.18$ & $0.02$ & $0.004$ & $0.005$ & $0.0015$ & $1\times10^{-4}$ & $1\times10^{-5}$ \\
 \hline
 $60 \ keV$ & 1.2 & 0.23 & 0.06 & 0.02 & 0.006 & $9\times10^{-4}$ & $9\times10^{-4}$ & $2\times10^{-4}$ & $3\times10^{-5}$ & $3\times 10^{-6}$ \\
 \hline
 $80 \ keV$ & $0.50$ & $0.10$ & $0.02$ & $0.007$ & $0.002$ & $0.0003$ & $2\times10^{-4}$ & $6\times10^{-5}$ & $1\times10^{-5}$ & $1\times10^{-6}$ \\
 \hline
		\end{tabular}
	\caption{Cross section results (in units of $10^{-18}$ cm$^2$) for charge transfer transitions into $n = 2,\ 3$ and $4$ levels.}
	\label{Tab:XsTrans}
\end{table*}

\section{Astrophysical Applications}
\label{Sec:Applications}

\subsection{Errors Associated with Extrapolating from Cross Sections with Lower $n$-values}

\begin{figure}
\centering
\includegraphics[width=3.4in]{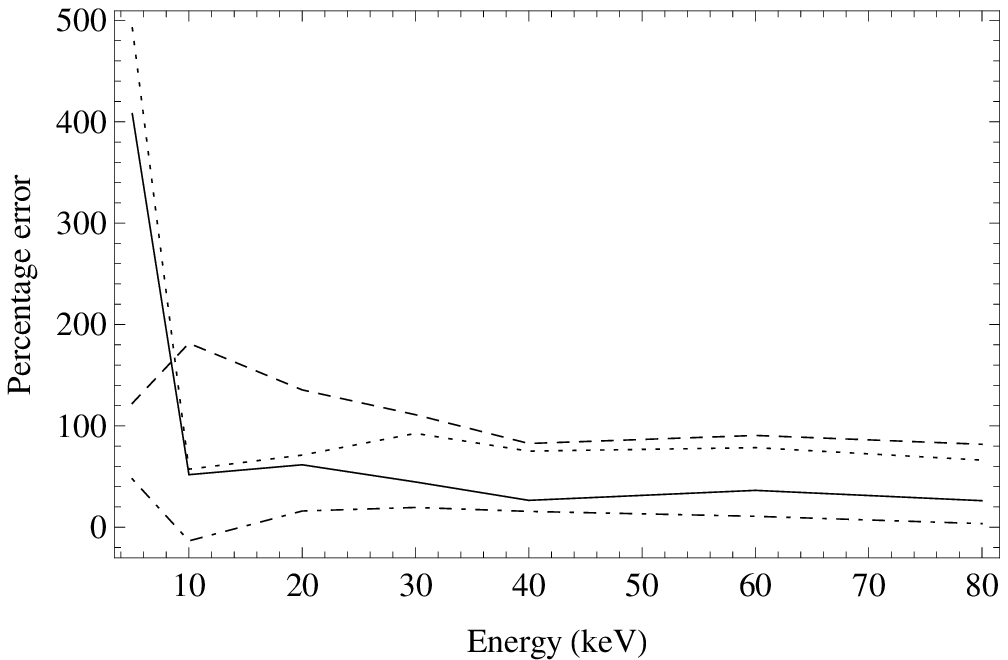}
\includegraphics[width=3.4in]{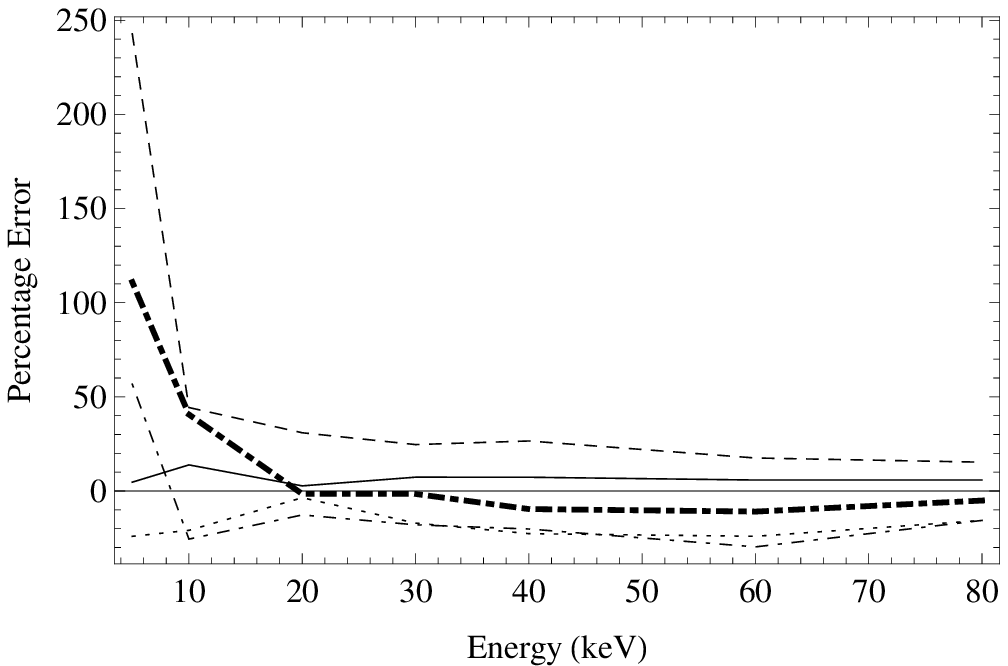}
\caption{\label{Fig:scaling} Errors associated with the calculated versus extrapolated (via equation~\ref{eq:scaling}) cross sections as a function of the impact energy. Upper panel shows percentage errors for $3s$ (solid curve), $3p_0$ (dashed curve), $3p_1$ (dotted curve), and $4s$ (dot-dashed curve) excitations. Lower panel shows errors for $4p_0$ (solid curve), $4p_1$ (dashed curve), $4d_0$ (dotted curve), $4d_1$ (dot-dashed curve), and $4d_2$ (thick dot-dashed curve) transitions.}
\end{figure}

In the absence of available cross sections, one is often forced to extrapolate from known cross sections with lower principal quantum numbers $n$ (e.g., \citealt{hm07}), thus generating errors in the calculation of line profiles and intensities which are unquantifiable.  With the benefit of now being able to calculate excitation cross sections for the H-H$^+$ collisional system, we quantify the error associated with using the scaling law in equation (\ref{eq:scaling}) for obtaining cross sections with $n \ge 3$.  In Figure \ref{Fig:scaling}, we see that the errors associated with extrapolating for $n=4$ cross sections from $n_0=3$ are typically a factor $\sim 2$.  The less necessary extrapolation of obtaining $n=3$ cross sections from $n_0=2$ (since data for $n=3$ is available) results in errors of a factor $\sim 5$.  Since these errors are non-negligible, they probably dominate any uncertainty associated with a numerical integration technique used to compute line profiles and intensities.  A direct calculation, such as the one we have performed in this study, is necessary in order to obtain accurate cross sections and in turn perform a spectral analysis of lines such as H$\beta$ at energies $\gtrsim 5$ keV.

\subsection{The Balmer Decrement}

\begin{figure}
\centering
\includegraphics[width=3.4in]{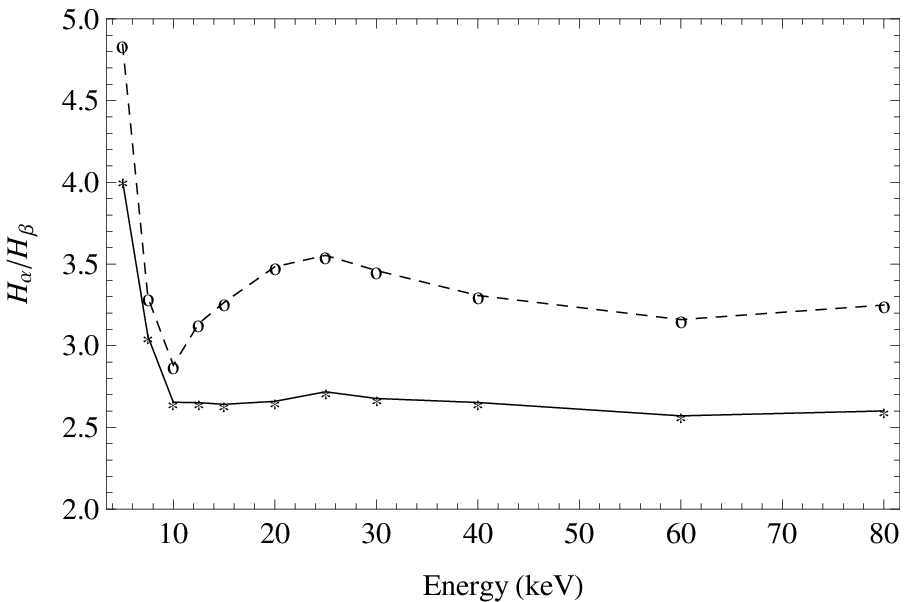}
\caption{\label{Fig:BalmDec} The Balmer decrement (H$\alpha$/H$\beta$) as a function of the impact energy for collisional excitation of hydrogen atoms by protons. Solid line corresponds to Case B, while dashed line shows Case A.}
\end{figure}

The Balmer decrement generally refers to the ratio of different lines in the Balmer series: H$\alpha$/H$\beta$, H$\beta$/H$\gamma$, etc.  It is somewhat insensitive to the electron temperature and (low) densities.  For example, H$\alpha$/H$\beta$ has a value of about 2--3 \citep{osterbrock68}, unless collisional excitation of hydrogen atoms by electrons dominate in which case its value is as high as 8 \citep{ap74}, which only occurs at electron densities $\sim 10^4$ cm$^{-3}$ or higher.  The relative insensitivity of the Balmer decrement to the atomic processes makes it an appropriate diagnostic for the presence of dust extinction, since the bluer lines in the Balmer series (e.g., H$\beta$) are subjected to increased extinction and thus the Balmer decrement attains a value larger than, e.g., 2--3 in the case of H$\alpha$/H$\beta$.  A high value of the Balmer decrement may also be attained if the population of excited hydrogen (e.g., 2s) is sufficient to cause self-absorption in the Balmer lines \citep{capriotti64, netzer75}.  As examples, the Balmer decrement has been used as a diagnostic in the study of supernovae (e.g., \citealt{aldering06}), active galactic nuclei (e.g., \citealt{Dong08}) and the \textit{Sloan Digital Sky Survey (SDSS)} sample of galaxies (e.g., \citealt{Groves11}).

Since our work was originally motivated by the study of fast astrophysical shocks, we use them as an example in the calculations presented in this sub-section.  In particular, we use the example of Balmer-dominated shocks, which are fast ($\sim 1000$ km s$^{-1}$) shocks impinging upon tenuous media ($\sim 1$ cm$^{-3}$).  As Balmer-dominated shocks are mostly observed around young ($\lesssim 1000$ years), Galactic supernova remnants, the production of photons via radiative recombination is unimportant since the recombination time is $\sim 10^4$ years.  The calculation of line intensities then requires knowledge of how the various atomic levels of hydrogen are populated via collisions as well as their subsequent rates of radiative decay.  For a strong shock (Mach number greatly exceeding unity), the relative velocity between the electrons/protons and hydrogen atoms is $\delta v = 3 v_s/4$, where $v_s$ denotes the shock velocity.  Thus, the shock velocity can be related to the interaction energy $E_{col}$ via
\begin{equation}
v_s = \frac{4}{3} \sqrt{ \frac{2 E_{col}}{m_{\rm H}} } \approx 1300 \mbox{ km s}^{-1} \left( \frac{E_{col}}{5 \mbox{ keV}} \right)^{1/2},
\end{equation}
where $m_{\rm H}$ is the mass of the hydrogen atom.  The cross sections presented in the present study are thus relevant to shocks with $v_s \approx 1300$--5200 km s$^{-1}$.

Formally, evaluating the rate at which a given $nl$ level is populated by collisions requires the calculation of the rate coefficient \citep{hm07},
\begin{equation}
{\cal R} = \int ~F_1\left(\vec{v}_1\right) ~F_2\left(\vec{v}_2\right) ~\Delta v ~\sigma\left(\Delta v \right) ~d^3v_1 ~d^3v_2.
\end{equation}
The preceding, six-dimensional integral is evaluated over all of the relative velocities ($\Delta v$) between the velocity distributions of the atoms ($F_1$) and electrons/protons ($F_2$), weighted by the relevant cross section ($\sigma$) of the atomic process being considered.

To gain an intuition for the functional dependence of the Balmer decrement on the interaction energy $E_{col}$, it is sufficient to consider either individual pairs of particles or particles in Delta-function distributions.  In this case, the rate coefficient reduces to ${\cal R} = \sigma \Delta v$.  The Balmer decrement is then the sum of the cross sections for the collisional population of each $nl$ level weighted by the appropriate branching ratio,
\begin{equation}
\mbox{H}\alpha/\mbox{H}\beta = \frac{ \sigma\left(\mbox{3s}\right) + B_{\mbox{3p,2s}} \sigma\left(\mbox{3p}\right) + \sigma\left(\mbox{3d}\right) }{B_{\mbox{4s,2p}} \sigma\left(\mbox{4s}\right) + B_{\mbox{4p,2s}} \sigma\left(\mbox{4p}\right) + B_{\mbox{4d,2p}} \sigma\left(\mbox{4d}\right) }.
\label{eq:decrement}
\end{equation}
The branching ratio is simply the Einstein A-coefficient for a given transition normalized by the Einstein A-coefficients of all of the transitions allowed by the electric dipole selection rule.  For example, $B_{\mbox{3p,2s}} = A_{\mbox{3p,2s}}/(A_{\mbox{3p,2s}} + A_{\mbox{3p,1s}})$.  

Two extreme limiting cases are typically considered: Case A and Case B \citep{Baker38, Seaton60, Osterbrock89}.  Case B occurs when the neutral hydrogen column density is large enough for Lyman lines to be optically thick. They undergo multiple scatterings and are eventually degraded into a Balmer line and Ly$\alpha$ or two-photon emission. Case A occurs when the neutral hydrogen column density is small enough for Lyman lines to be optically thin and so freely escape the cloud. In a Case A scenario, we have (e.g., \citealt{hs08}),
\begin{equation}
\begin{split}
&B_{\mbox{3p,2s}} \approx 0.1183,\\
&B_{\mbox{4s,2p}} \approx 0.5841,\\
&B_{\mbox{4p,2s}} \approx 0.1191,\\
&B_{\mbox{4d,2p}} \approx 0.7456.\\
\end{split}
\end{equation}
In a Case B scenario, these branching ratios are essentially unity, since $B_{\mbox{4p,3s}} \sim 10^{-2}$ and $B_{\mbox{4p,3s}} \sim 10^{-3}$.

In Figure \ref{Fig:BalmDec}, we show calculations of the Balmer decrement, for both Case A and B, using equation (\ref{eq:decrement}).  For simplicity, we consider only the collisional excitation (and not charge transfer) of hydrogen atoms by protons; collisional excitation by electrons is sub-dominant at these energies.  It is apparent that H$\alpha$/H$\beta$ remains somewhat constant at values of 2--3 at high energies and starts to grow rapidly at energies below $\sim 10$ keV. At 5 keV, it reaches values of 4--5.  If we use the values of 2--3 as a baseline, then this corresponds to the true dust extinction being over-estimated by $\Delta A_V=$1--3 (based on $R_V = 3.1$ model and extinction curves from~\cite{Weingartner01}).  Thus, we caution the use of the Balmer decrement as a diagnostic for dust extinction as it possesses some sensitivity to the atomic physics at energies $\lesssim 10$ keV.

\section{Summary}
\label{Sec:Summary}

In this work, we introduced a new formalism for computing precise cross-sections for high-$nl$ proton-hydrogen collisions and developed a numerical code which implements our formalism. We further used our code to obtain accurate cross sections for collisions between protons and hydrogen atoms which start in the ground state. Our computed cross sections focused on the energy range of direct interest for the studies of Balmer-dominated shocks; as the observed spectra of these shocks improve in quality and precision, our cross sections are required for doing a detailed interpretation, e.g., to estimate the degree to which electron and proton temperature are equilibrated.

The code, \texttt{BDSCx}, introduced in this paper has a large number of potential applications in atomic physics. Using the $q$-drop procedure and curvilinear coordinates proposed in our work enables relatively inexpensive calculations of charge transfer and excitation cross sections for proton-hydrogen collision with $n \lesssim 7$. In this paper we focused on the formalism and numerical implementation of the proposed cross section calculations, while detailed tests of the $q$-drop procedure and the cross section results for $n > 4$ will be reported in future papers. 

\section*{Acknowledgements}

We are grateful to Avi Loeb for his help with getting the time on the Odyssey cluster supported by the FAS Science Division Research Computing Group at Harvard University. We are also grateful to Mark Scheel for helping to get the time on the SHC cluster at Caltech. D.T. and C.H. are supported by the U.S. Department of Energy (DE-FG03-92-ER40701) and the National Science Foundation (AST-0807337). C.H. is supported by the David and Lucile Packard Foundation.  K.H. is supported by the Zwicky Prize Fellowship of the Institute for Astronomy of ETH Z\"{u}rich.

\clearpage
\pagebreak[4]

\appendix

\section{Fitting Functions for Computed Cross Sections}

In order to facilitate the broader use of our cross sections, we are providing fitting functions which allow a more straightforward utilization of the obtained results. We fit the obtained cross sections for excitations and charge transfer into $3s$, $3p$, $3d$, $4s$, $4p$, $4d$, and $4f$ states with a series of Chebyshev orthogonal polynomials. The fitting function is of the form:
\begin{equation}
{\cal F}\left(x;\vec{A}\right) = \exp{\left(\frac{A_0}{2} + \sum^{4}_{i=1} ~A_i ~{\cal C}_i\left(x\right)\right)},
\end{equation}
where the coefficients $\vec{A}=A_i$ for $0 \le i \le 7$ are the fitting parameters.  The quantities ${\cal C}_i$ are the Chebyshev orthogonal polynomials:
\begin{eqnarray}
{\cal C}_1\left(x\right) = x,\\
{\cal C}_2\left(x\right) = 2x^2 - 1,\\
{\cal C}_3\left(x\right) = 4x^3 - 3x,\\
{\cal C}_4\left(x\right) = 8\left(x^4 - x^2 \right) + 1,\\
{\cal C}_5\left(x\right) = 16x^5 - 20x^3 + 5x,\\
{\cal C}_6\left(x\right) = 32x^6 - 48x^4 + 18x^2 - 1,\\
{\cal C}_7\left(x\right) = 64x^7 - 112x^5 + 56x^3 - 7x.
\end{eqnarray}
The fitting variable $x$ is defined as
\begin{equation}
x= \frac{ \ln{\left(E_{col}/E_{\rm{min}}\right)} - \ln{\left(E_{\rm{max}}/E_{col}\right)} }{\ln{\left(E_{\rm{max}}/E_{\rm{min}}\right)}},
\end{equation}
where $E_{col}$ is the relative energy between the proton and hydrogen atom; $E_{\rm{min}} = 5$ keV and $E_{\rm{max}} = 80$ keV are the respective minimum and maximum energies in our simulation.  We use the Newton fitting algorithm which provides a local-optimal fit to the array of available cross section results. 

The fitting parameters are provided in Tables~\ref{Tab:Fit} and~\ref{Tab:Fitn}. We note, that for purposes of precise analysis the use of actual cross section data points provided in Tables~\ref{Tab:Xs} and~\ref{Tab:XsTrans} and simple spline extrapolation will produce more accurate results.

\begin{table*}
	\centering
		\begin{tabular}{ | l | c | c | c | c | c | c | c | r | }
		\hline
 \text{} & \text{A0} & \text{A1} & \text{A2} & \text{A3} & \text{A4} & \text{A5} & \text{A6} & \text{A7} \\
 \hline
 \hline
 \text{3s} & 0.811 & 1.04 & -0.369 & -0.173 & -0.00396 & 0.152 & -0.0638 & 0.00303 \\
 \hline
 \text{3p} & 3.35 & 1.03 & -0.0785 & -0.0904 & -0.0665 & 0.064 & $4\times 10^{-7}$ & -0.00166 \\
 \hline
 \text{3d} & 1.59 & -0.0412 & 0.0595 & -0.279 & -0.00958 & 0.105 & -0.0475 & -0.0168 \\
 \hline
 \text{4s} & -1.16 & 1.08 & -0.371 & -0.0661 & -0.097 & 0.191 & -0.0702 & 0.015 \\
 \hline
 \text{4p} & 0.0355 & 0.187 & -0.108 & -0.186 & -0.0919 & 0.0756 & -0.00533 & -0.0405 \\
 \hline
 \text{4d} & 0.835 & 1.4 & -0.235 & -0.0475 & -0.0222 & 0.0115 & 0.0116 & -0.00492 \\
 \hline
 \text{4f} & -3.71 & -0.925 & -0.122 & -0.325 & -0.0885 & 0.108 & 0.016 & -0.0633 \\
 \hline
		\end{tabular}
	\caption{Fitting coefficients $A_i$ corresponding to excitation transitions into $n=3$ and $n=4$ states.}
	\label{Tab:Fit}
\end{table*}

\begin{table*}
	\centering
		\begin{tabular}{ | l | c | c | c | c | c | c | c | r | }
		\hline
 \text{} & \text{A0} & \text{A1} & \text{A2} & \text{A3} & \text{A4} & \text{A5} & \text{A6} & \text{A7} \\
 \hline
 \hline
 \text{3s} & 1.15 & 0.963 & -1.35 & -0.242 & 0.14 & 0.00596 & -0.0572 & 0.0564 \\
 \hline
 \text{3p} & 1.14 & -0.699 & -1.05 & -0.0922 & 0.0511 & -0.0818 & -0.054 & -0.04 \\
 \hline
 \text{3d} & -1.65 & -2.38 & -0.763 & 0.137 & -0.00275 & -0.0614 & 0.0162 & -0.0629 \\
 \hline
 \text{4s} & -0.798 & 1.14 & -1.24 & -0.321 & 0.173 & -0.0128 & -0.0195 & -0.0199 \\
 \hline
 \text{4p} & -2.78 & -2. & -1.09 & -0.208 & -0.127 & -0.136 & 0.0375 & 0 \\
 \hline
 \text{4d} & -0.926 & -0.344 & -1.13 & -0.164 & 0.135 & -0.0675 & -0.0372 & 0.00486 \\
 \hline
 \text{4f} & -6.95 & -3.49 & -0.617 & 0.297 & 0.174 & 0.0395 & -0.0364 & 0.000221 \\
 \hline
		\end{tabular}
	\caption{Fitting coefficients $A_i$ corresponding to charge transfer into $n=3$ and $n=4$ states.}
	\label{Tab:Fitn}
\end{table*}

\section{Cross Section Plots}

This Appendix provides plots of cross sections for excitation and charge transfer transitions into $3s$, $3p$, $3d$, $4s$, $4p$, $4d$, and $4f$ states. Our results are compared with the results of theoretical studies for $n = 3$ from~\cite{Kolakowska98} and~\cite{Winter09}. Along with the discrete cross section data points, we are providing the results of our Chebyshev polynomial fits described in Appendix A.

\begin{figure*}
\centering
\includegraphics[width=3.3in]{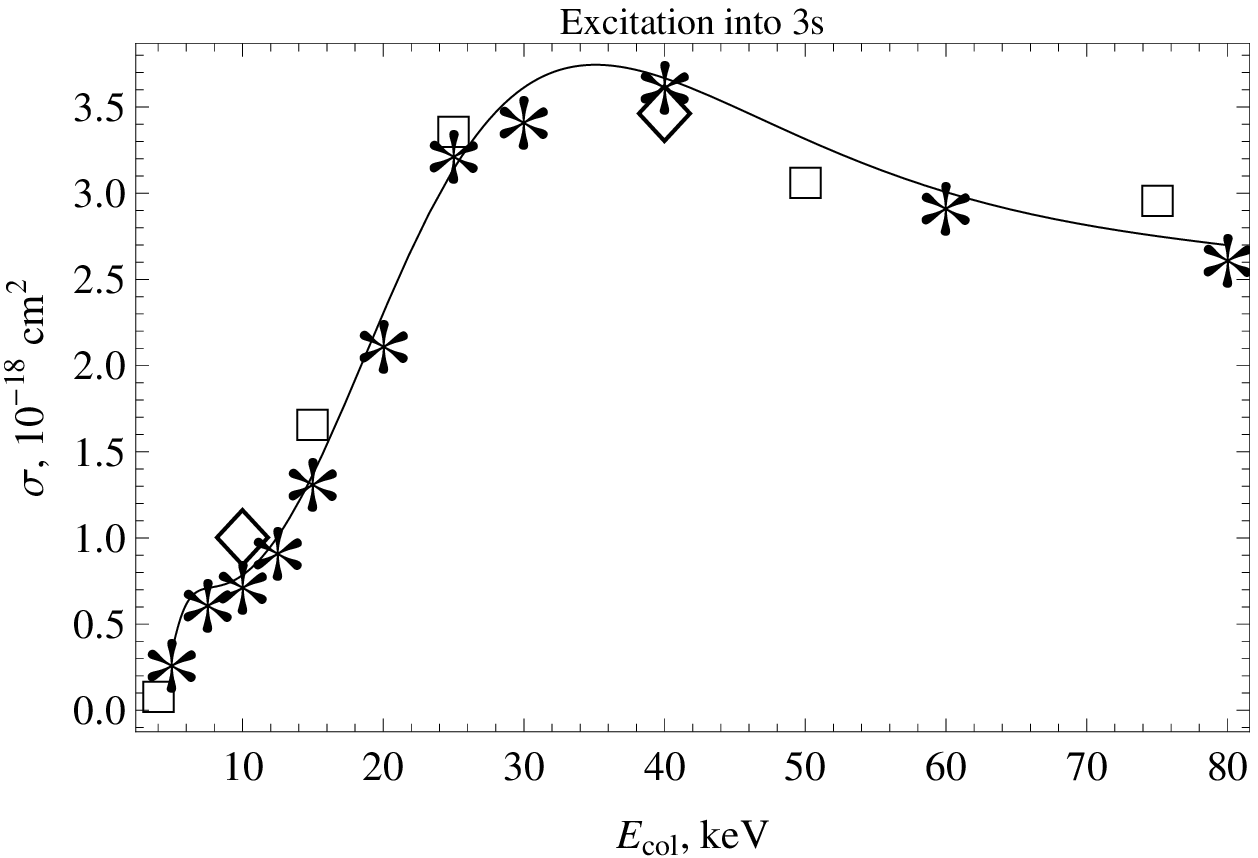}
\includegraphics[width=3.4in]{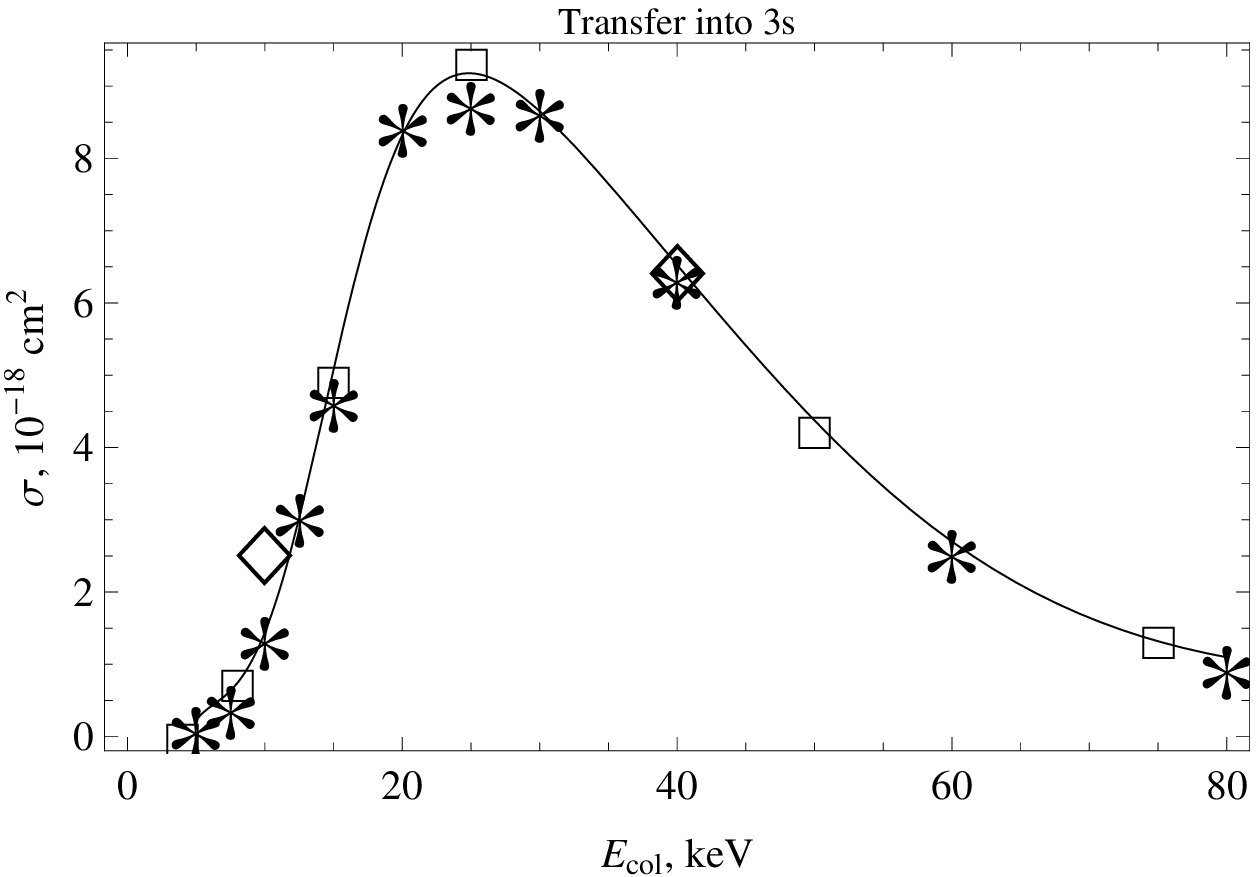}
\caption{\label{Fig:ResPlots3s} Cross sections for excitation (left panel) and charge transfer (right panel) into $3s$ state. Stars show results of our calculations, diamonds correspond to Kolakowska (1998), and squares show results from Winter (2009). Solid line corresponds to our Chebyshev polynomial fit.}
\end{figure*}

\begin{figure*}
\centering
\includegraphics[width=3.4in]{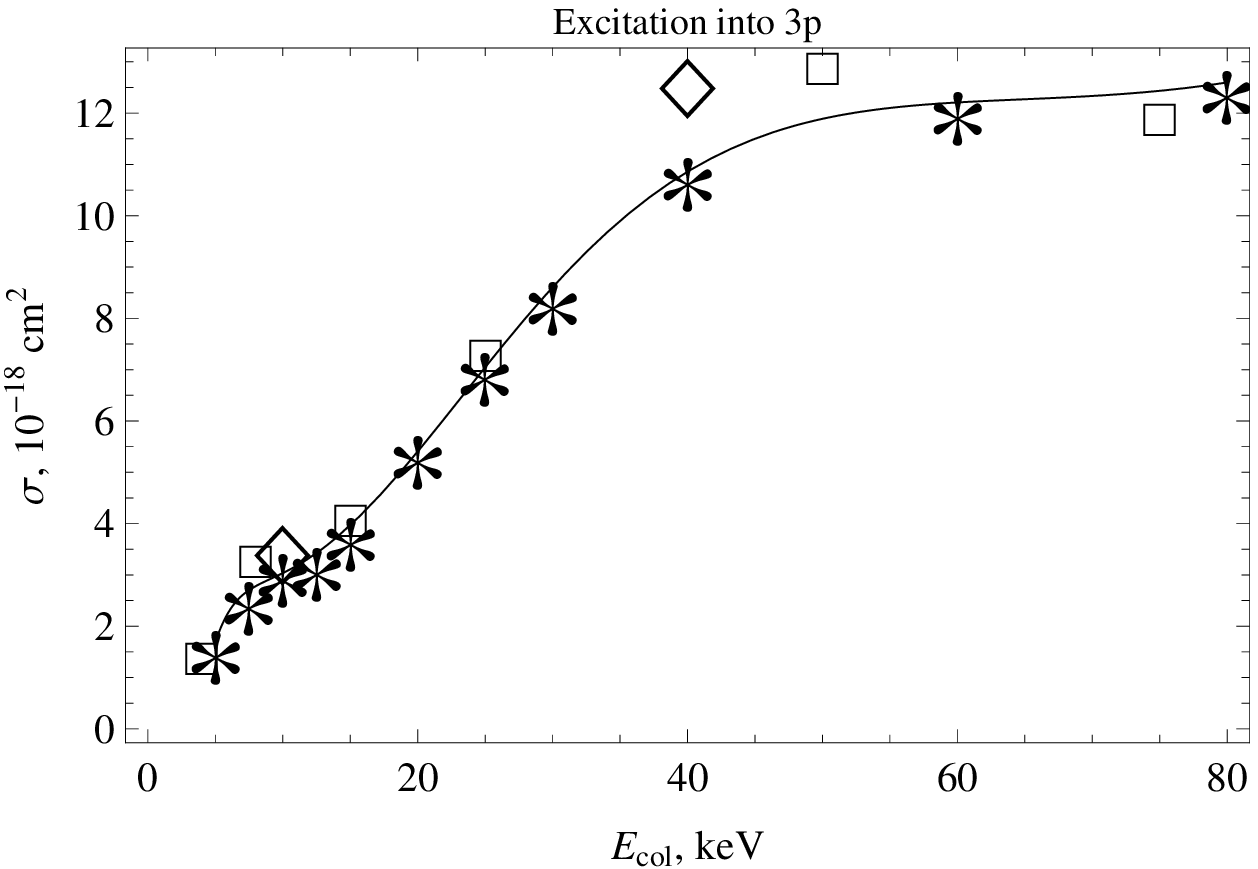}
\includegraphics[width=3.4in]{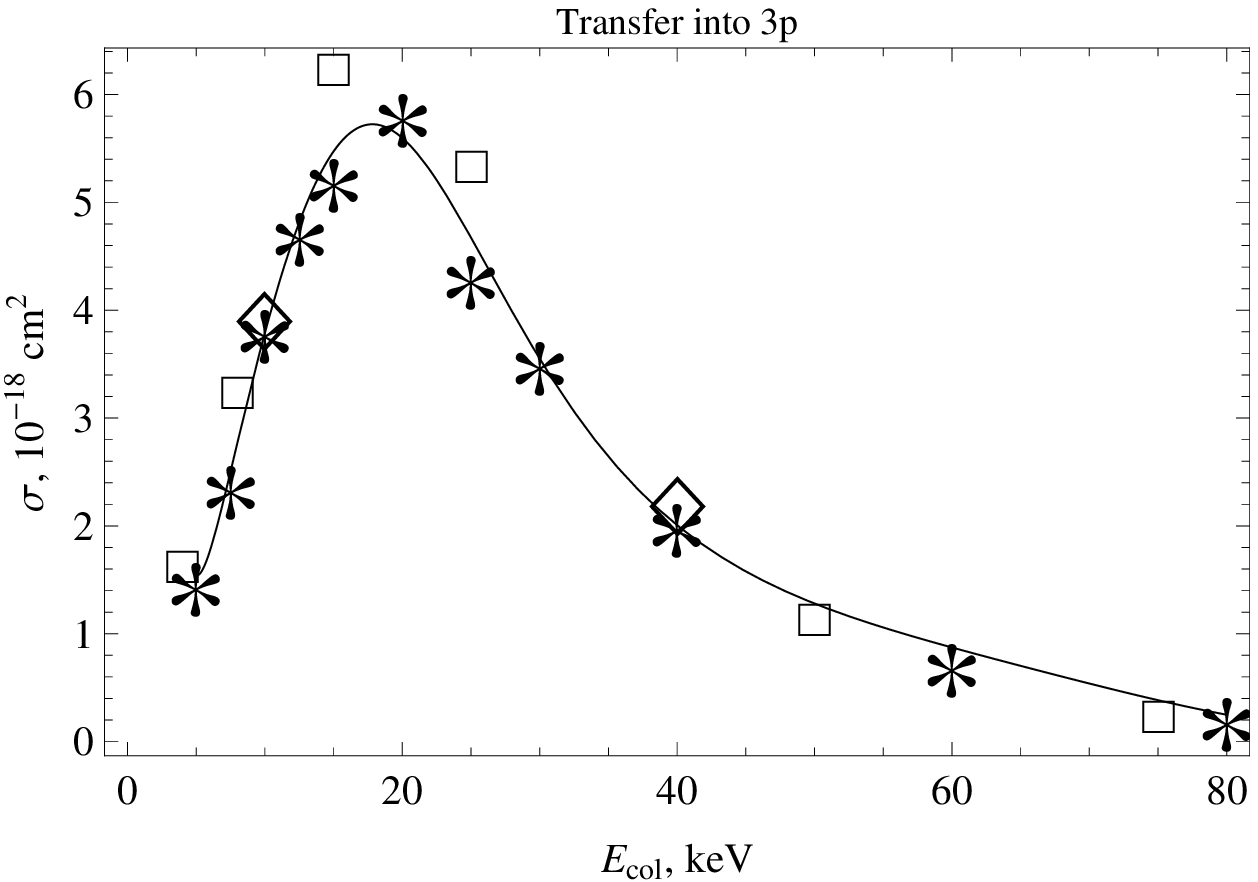}
\caption{\label{Fig:ResPlots3p} Cross sections for excitation (left panel) and charge transfer (right panel) into $3p$ state. Stars show results of our calculations, diamonds correspond to Kolakowska (1998), and squares show results from Winter (2009). Solid line corresponds to our Chebyshev polynomial fit.}
\end{figure*}

\begin{figure*}
\centering
\includegraphics[width=3.4in]{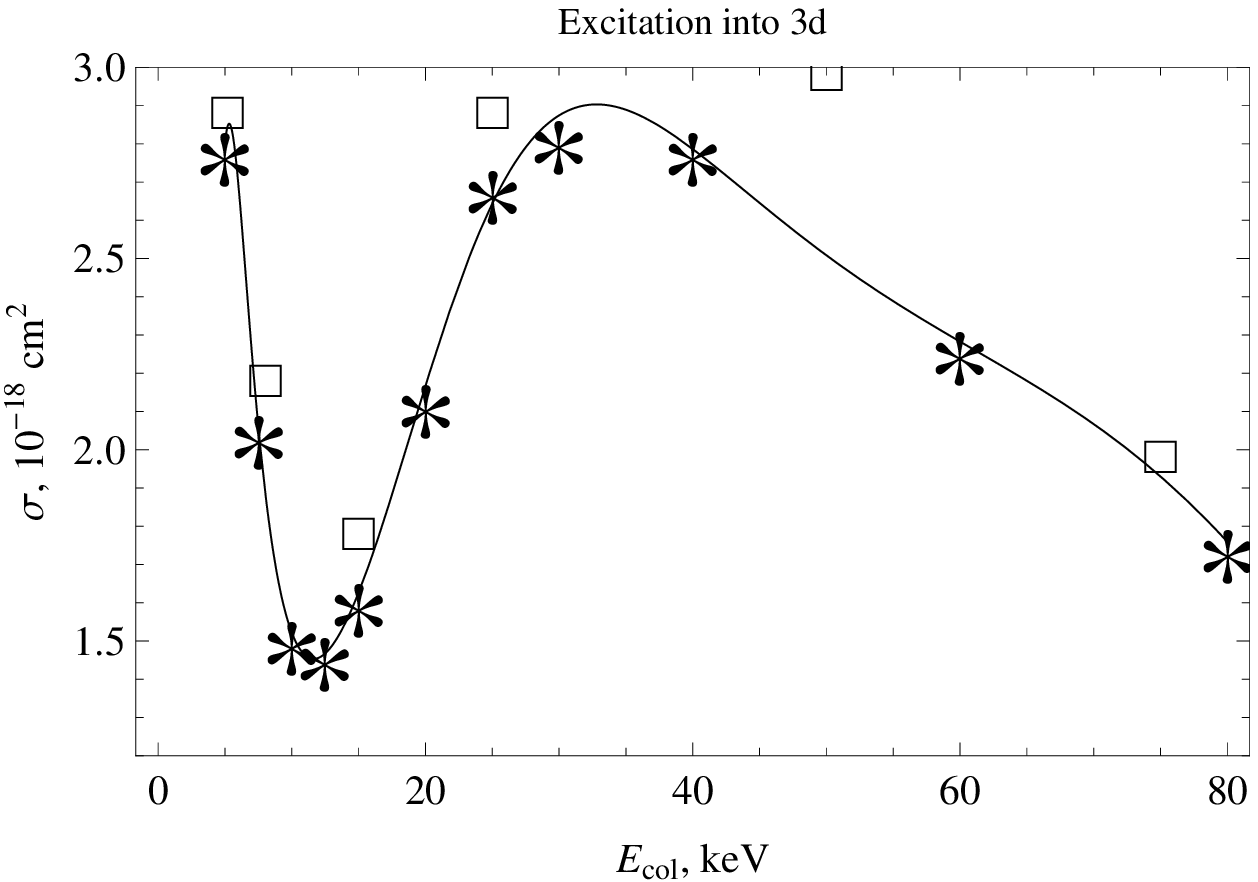}
\includegraphics[width=3.4in]{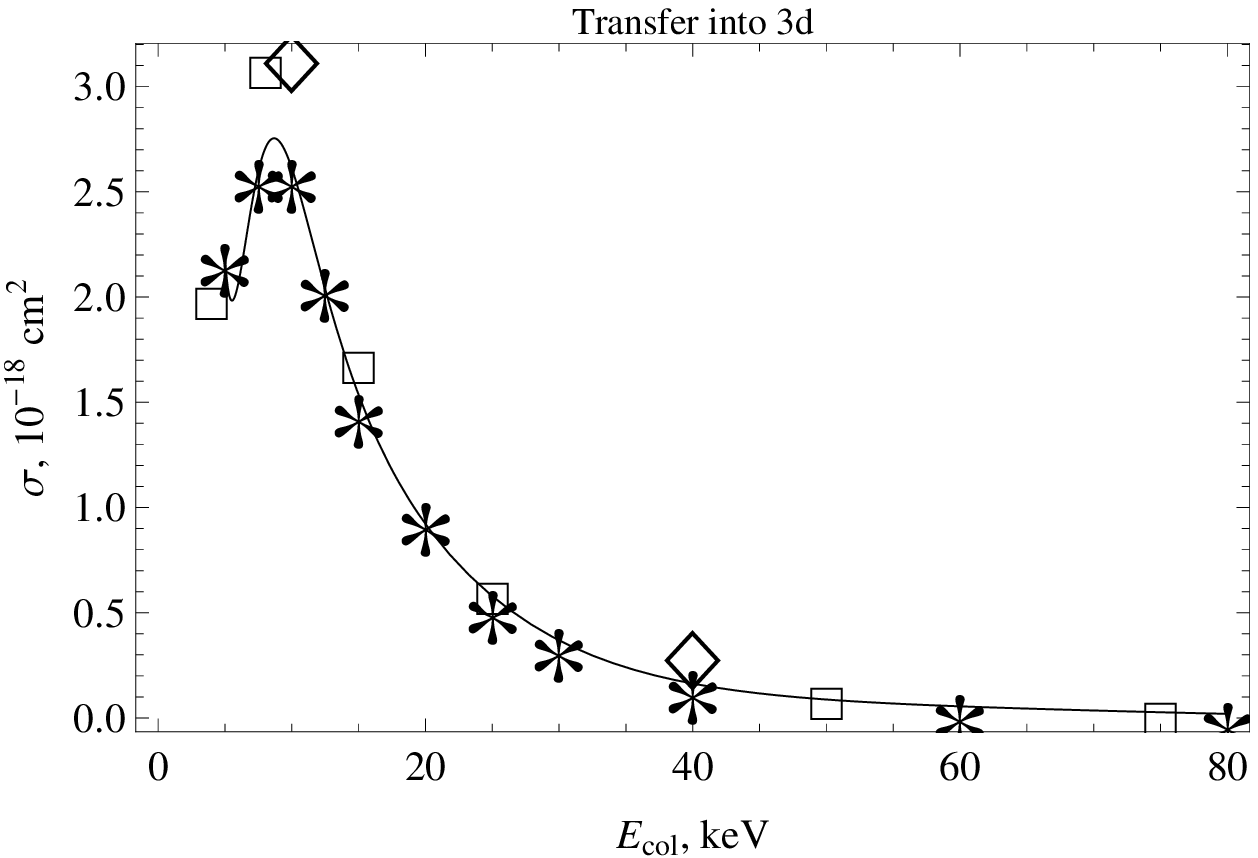}
\caption{\label{Fig:ResPlots3d} Cross sections for excitation (left panel) and charge transfer (right panel) into $3d$ state. Stars show results of our calculations, diamonds correspond to Kolakowska (1998), squares show results from Winter (2009). Solid line corresponds to our Chebyshev polynomial fit.}
\end{figure*}

\begin{figure*}
\centering
\includegraphics[width=3.4in]{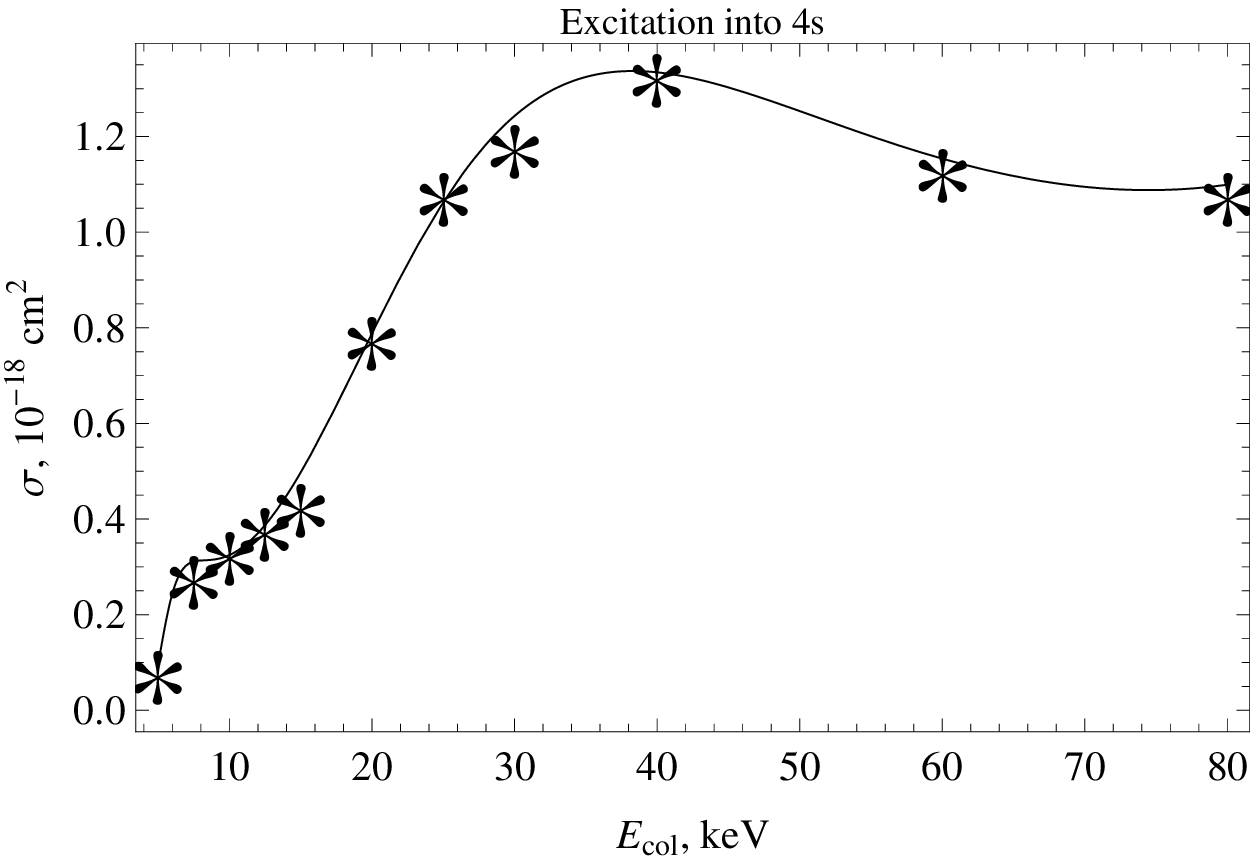}
\includegraphics[width=3.4in]{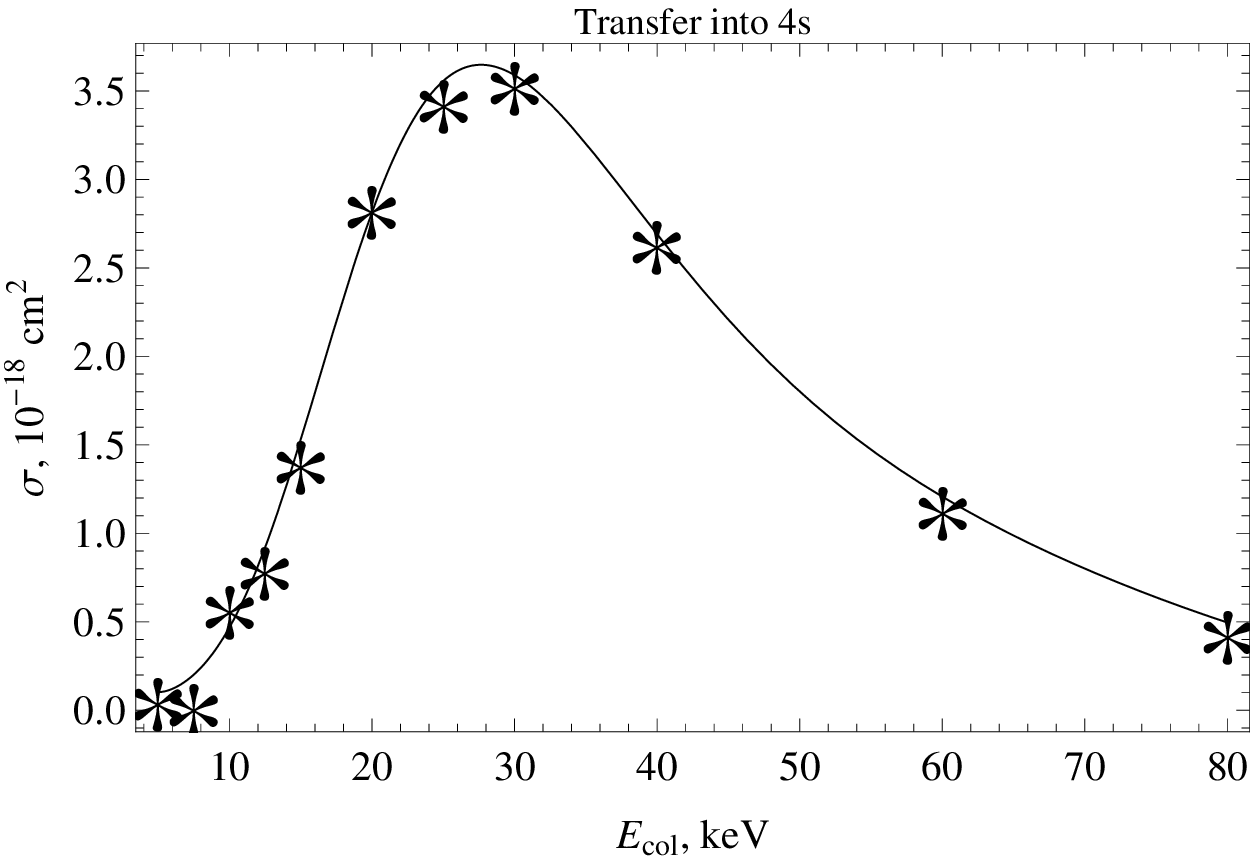}
\caption{\label{Fig:ResPlots4s} Cross sections for excitation (left panel) and charge transfer (right panel) into $4s$ state. Stars show results of our calculations. Solid line corresponds to our Chebyshev polynomial fit.}
\end{figure*}

\begin{figure*}
\centering
\includegraphics[width=3.4in]{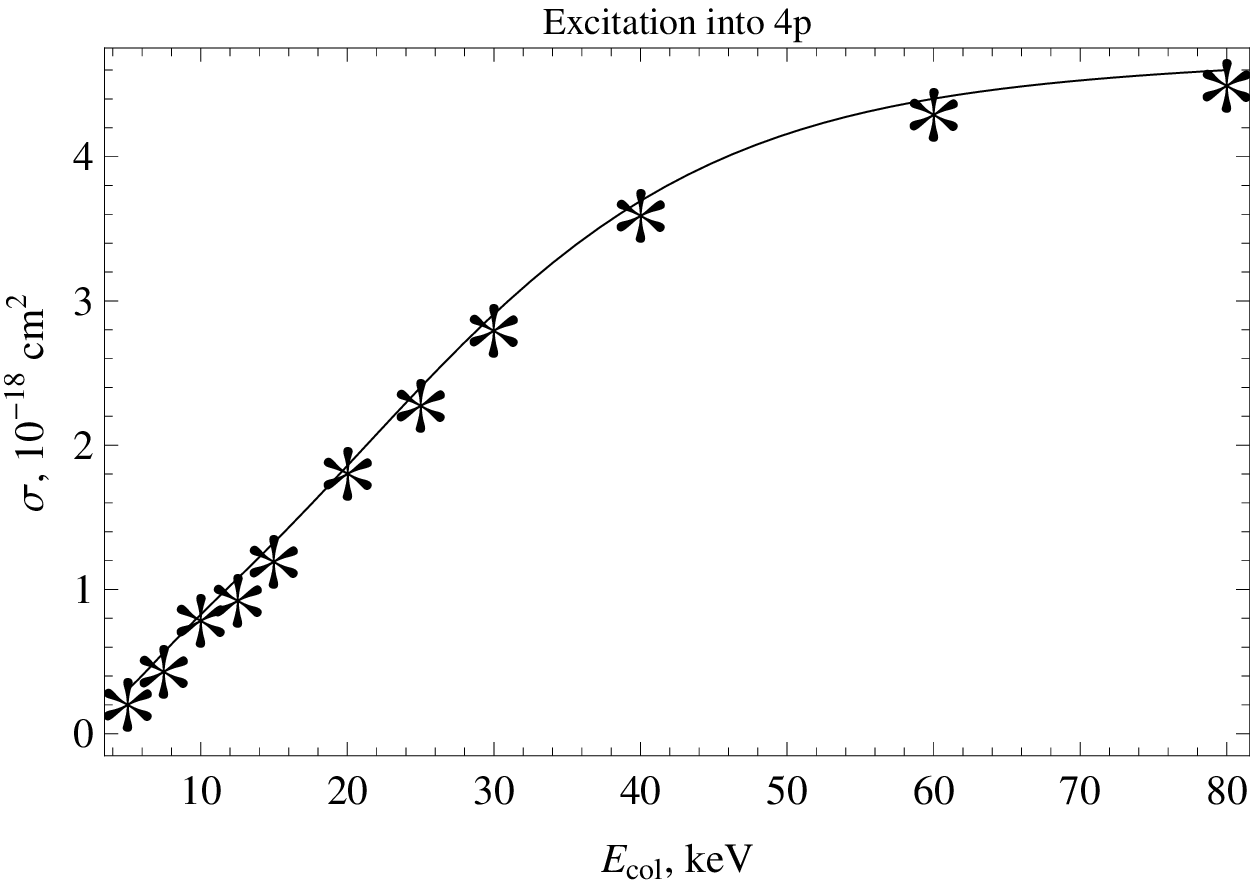}
\includegraphics[width=3.4in]{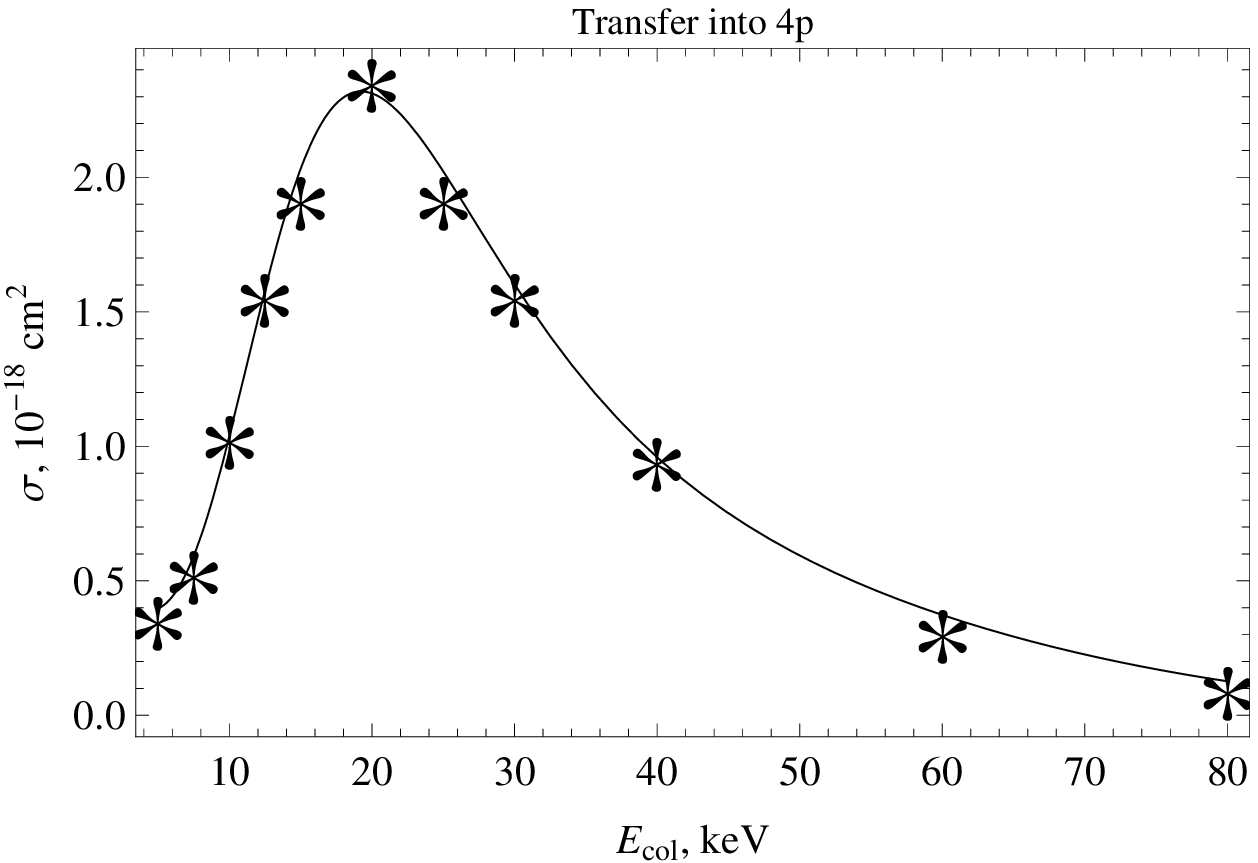}
\caption{\label{Fig:ResPlots4p} Cross sections for excitation (left panel) and charge transfer (right panel) into $4p$ state.  Stars show results of our calculations. Solid line corresponds to our Chebyshev polynomial fit.}
\end{figure*}

\begin{figure*}
\centering
\includegraphics[width=3.4in]{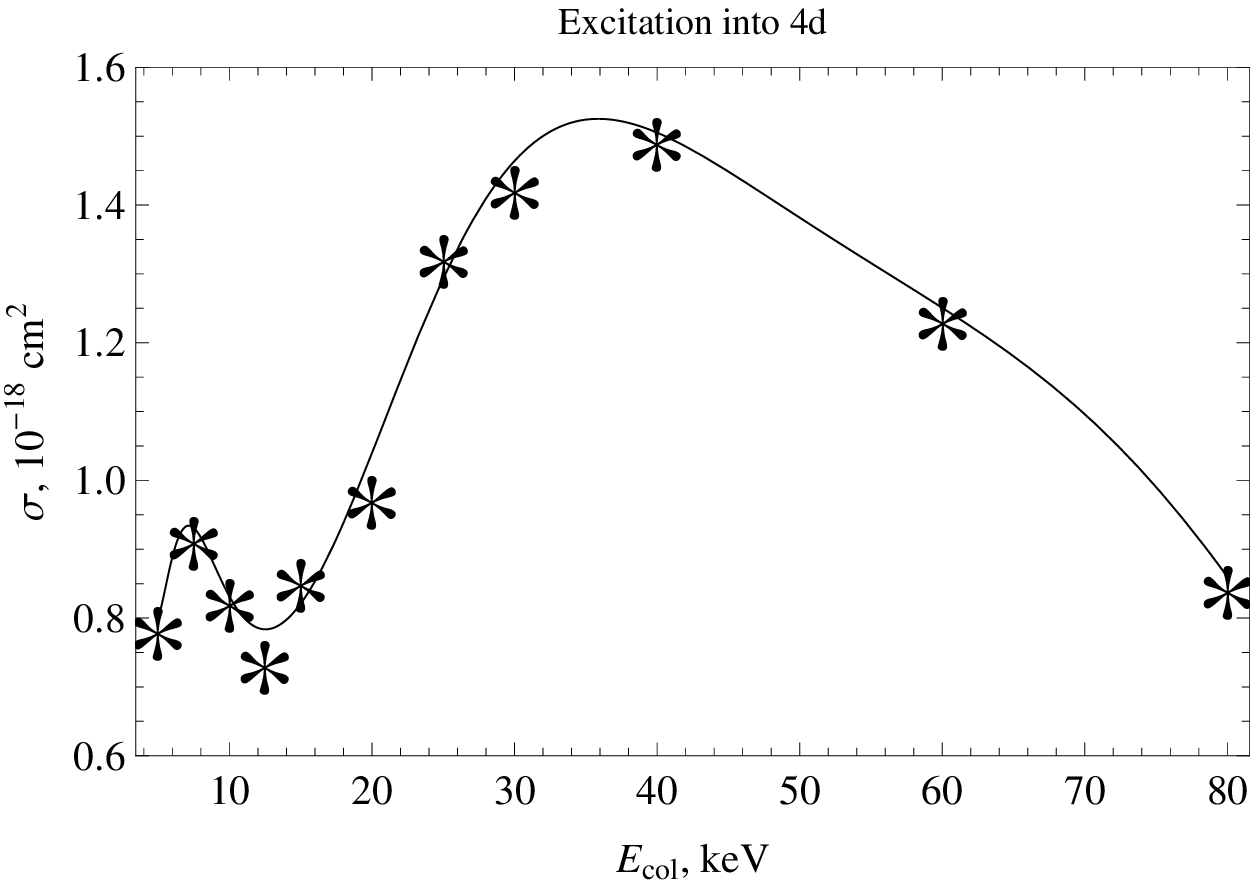}
\includegraphics[width=3.4in]{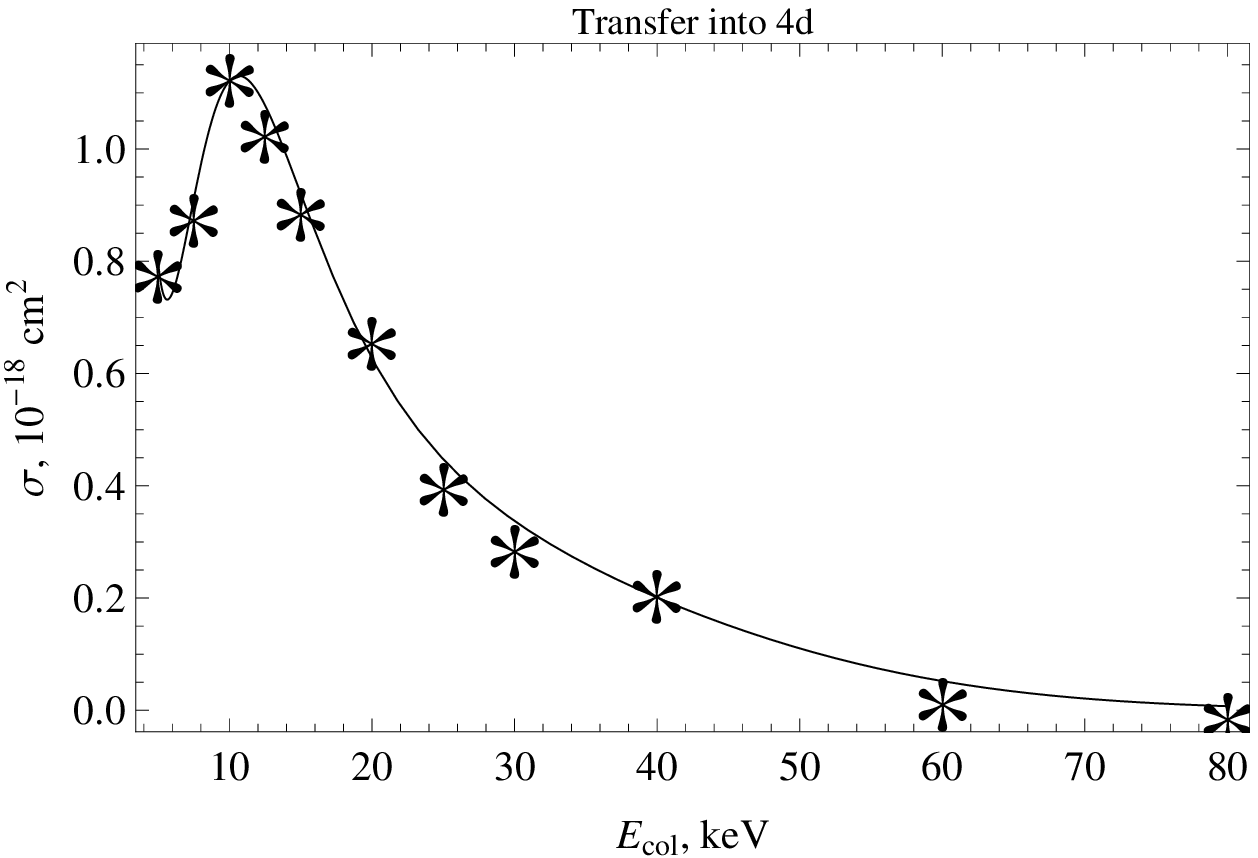}
\caption{\label{Fig:ResPlots4d} Cross sections for excitation (left panel) and charge transfer (right panel) into $4d$ state.  Stars show results of our calculations. Solid line corresponds to our Chebyshev polynomial fit.}
\end{figure*}

\begin{figure*}
\centering
\includegraphics[width=3.4in]{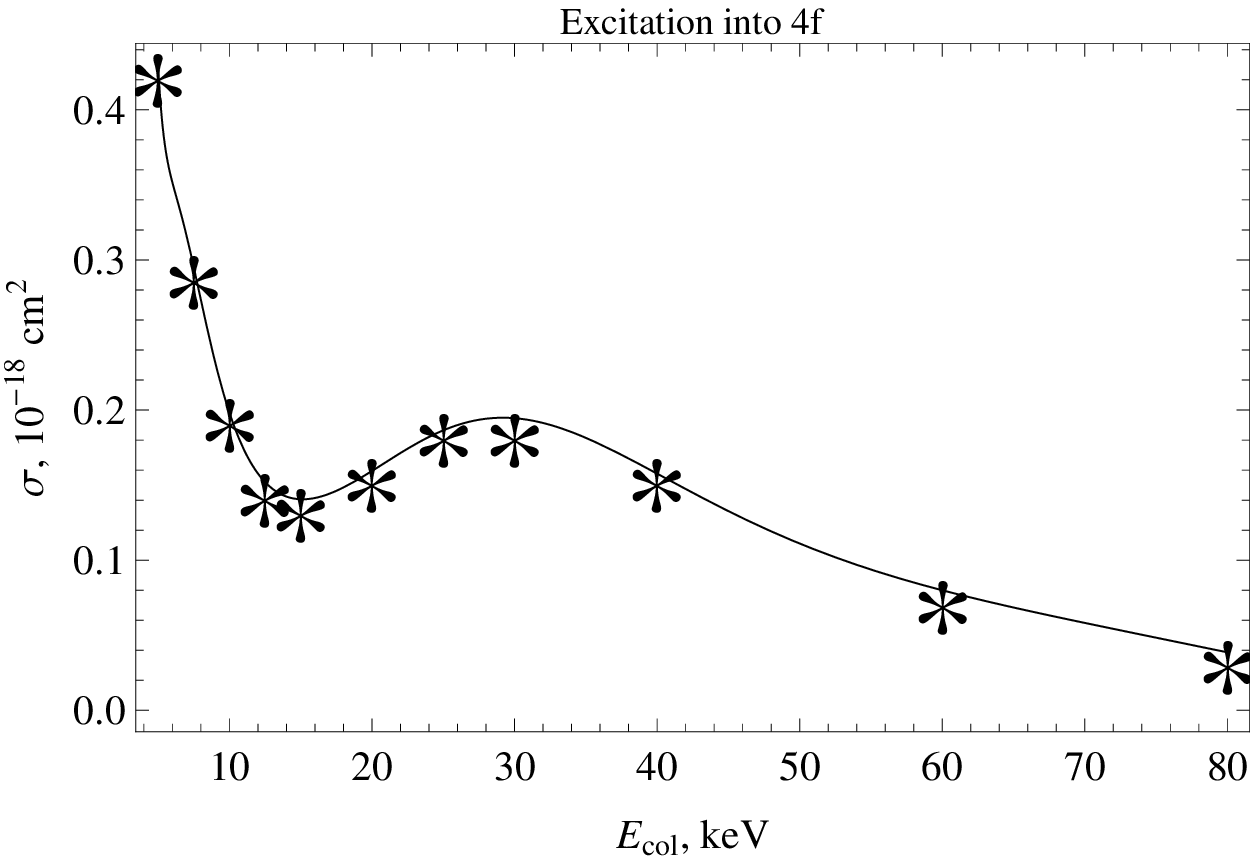}
\includegraphics[width=3.4in]{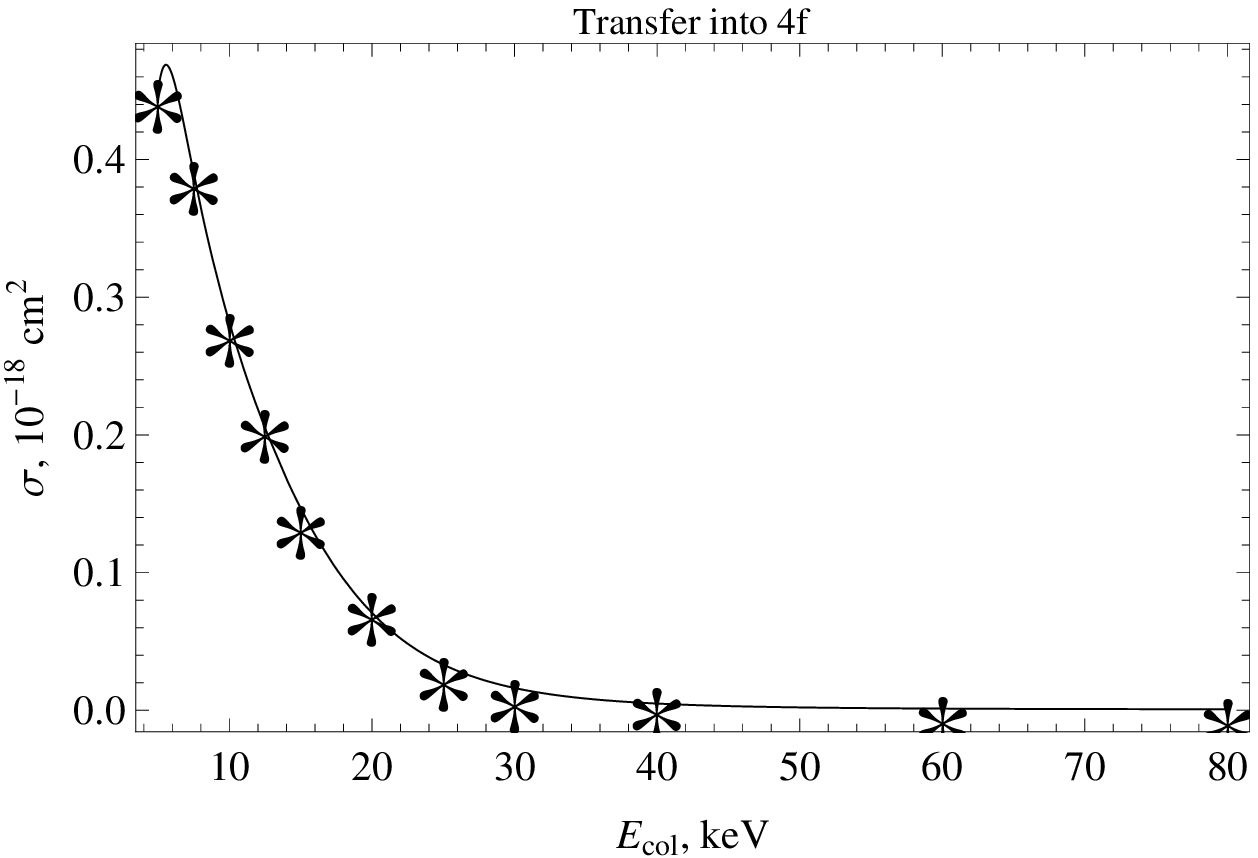}
\caption{\label{Fig:ResPlots4f} Cross sections for excitation (left panel) and charge transfer (right panel) into $4f$ state.  Stars show results of our calculations. Solid line corresponds to our Chebyshev polynomial fit.}
\end{figure*}

\end{document}